\documentclass[journal]{vgtc}                       

\ifpdf
  \pdfoutput=1\relax                   
  \usepackage{graphicx}                
  \DeclareGraphicsExtensions{.pdf,.png,.jpg,.jpeg} 
\else
  \usepackage{graphicx}                
  \DeclareGraphicsExtensions{.eps}     
\fi%

\graphicspath{{figures/}{pictures/}{images/}{./}} 

\usepackage{microtype}                 
\PassOptionsToPackage{warn}{textcomp}  
\usepackage{textcomp}                  
\usepackage{mathptmx}                  
\usepackage{times}                     
\usepackage{cite}                      
\usepackage{booktabs}                  
\usepackage{makecell}
\usepackage{xcolor}
\usepackage{amsmath}
\usepackage{amssymb}
\usepackage{mathtools}
\usepackage{bm}
\usepackage{algorithm}
\usepackage{algpseudocode}
\usepackage{array}
\newcolumntype{L}[1]{>{\raggedright\let\newline\\\arraybackslash\hspace{0pt}}m{#1}}
\newcolumntype{C}[1]{>{\centering\let\newline\\\arraybackslash\hspace{0pt}}m{#1}}
\newcolumntype{R}[1]{>{\raggedleft\let\newline\\\arraybackslash\hspace{0pt}}m{#1}}
\usepackage[caption=false,font=footnotesize,labelfont=sf,textfont=sf]{subfig}
\usepackage[percent]{overpic}
\usepackage[export]{adjustbox}

\usepackage[frozencache,cachedir=.]{minted}

\usepackage{multirow}
\usepackage[framemethod=TikZ]{mdframed}

\newcommand{\etAl}{\textit{et~al.}}
\newcommand{\pluseq}{\mathrel{+}=}
\newcommand{\R}{\mathbb{R}}

\newcommand{\dd}{\text{d}}

\algnewcommand\algorithmicparameters{\textbf{Parameters:}}
\algnewcommand\Parameters{\item[\algorithmicparameters]}
\algnewcommand\algorithmicinput{\textbf{Input:}}
\algnewcommand\Input{\item[\algorithmicinput]}
\algnewcommand\algorithmicoutput{\textbf{Output:}}
\algnewcommand\Output{\item[\algorithmicoutput]}
\algnewcommand\algorithmicforward{\textbf{Forward:}}
\algnewcommand\Forward{\item[\algorithmicforward]}
\algnewcommand\algorithmicbackward{\textbf{Backward:}}
\algnewcommand\Backward{\item[\algorithmicbackward]}

\def\orcidbase{https://orcid.org/}
\newcommand{\orcid}[1]{\hbox{\href{\orcidbase #1}{\includegraphics{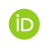}}}}

\onlineid{1136}

\vgtccategory{research}
\vgtcpapertype{algorithm/technique}

\title{Differentiable Direct Volume Rendering}

\author{Sebastian Weiss\orcid{0000-0003-4399-3180}\thanks{e-mail: sebastian13.weiss@tum.de}~
and Rüdiger Westermann\orcid{0000-0002-3394-0731}\thanks{e-mail: westermann@tum.de}}
\affiliation{\scriptsize Technical University of Munich}
\shortauthortitle{Weiss \MakeLowercase{\textit{et al.}}: Differentiable Direct Volume Rendering}

\teaser{
 \includegraphics[width=\linewidth,trim=0 5 0 0,clip]{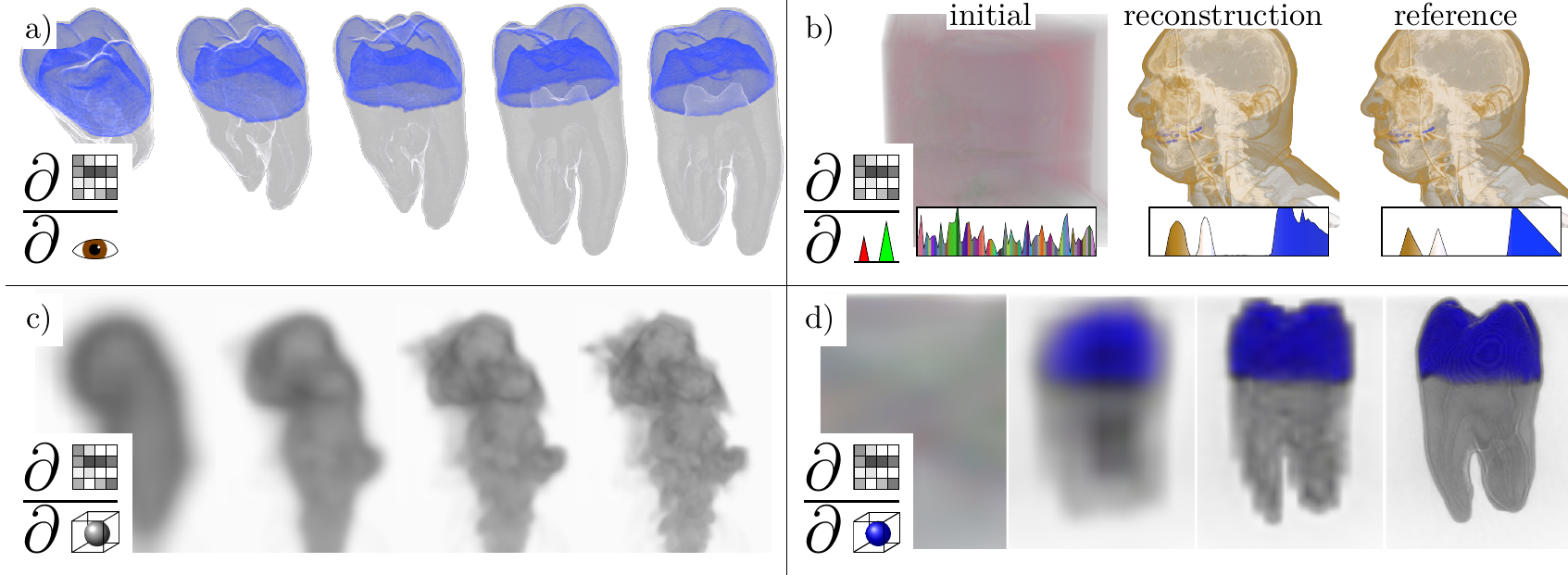}\vspace*{-2mm}%
 \caption{A fully differentiable direct volume renderer is used for a) viewpoint optimization, b) transfer function optimization, and optimization of voxel properties using c) an absorption-only model and d) an emission-absorption model with rgb$\alpha$ transfer functions. a), c) and d) show intermediate results of the optimization process until convergence.}
 \label{fig:teaser}
}

\abstract{
We present a differentiable volume rendering solution that provides differentiability of all continuous parameters of the volume rendering process. This differentiable renderer is used to steer the parameters towards a setting with an optimal solution of a problem-specific objective function. We have tailored the approach to volume rendering by enforcing a constant memory footprint via analytic inversion of the blending functions. This makes it independent of the number of sampling steps through the volume and facilitates the consideration of small-scale changes. The approach forms the basis for automatic optimizations regarding external parameters of the rendering process and the volumetric density field itself. We demonstrate its use for automatic viewpoint selection using differentiable entropy as objective, and for optimizing a transfer function from rendered images of a given volume. Optimization of per-voxel densities is addressed in two different ways: First, we mimic inverse tomography and optimize a 3D density field from images using an absorption model. This simplification enables comparisons with algebraic reconstruction techniques and state-of-the-art differentiable path tracers. Second, we introduce a novel approach for tomographic reconstruction from images using an emission-absorption model with post-shading via an arbitrary transfer function.
} 

\keywords{Differentiable rendering, Direct Volume Rendering, Automatic Differentiation}

\vgtcinsertpkg

\begin{document}


\firstsection{Introduction}

\maketitle

Differentiable direct volume rendering (DiffDVR) can serve as a basis for a multitude of automatic optimizations regarding external parameters of the rendering process such as the camera, the transfer function (TF), and the integration stepsize, as well as the volumetric scalar field itself. DiffDVR computes derivatives of the rendered pixel values with respect to these parameters and uses these derivatives to steer the parameters towards an optimal solution of a problem-specific objective (or loss) function. DiffDVR is in particular required when using neural network-based learning tasks, where derivatives need to be propagated seamlessly through the network for training end-to-end regarding the loss function. 

While a number of approaches have been proposed for differentiable surface rendering~\cite{kato2020differentiable}, approaches focusing on differentiable rendering in the context of volume visualization are rare. For surface rendering, one objective is on the optimization of scene parameters like material properties, lighting conditions, or even geometric shape, to achieve matchings of synthetic and real images in computer vision tasks. Others have used implicit surface representations encoded via volumetric signed distance functions to derive analytic gradients for image-based shape reconstruction tasks~\cite{liu2019learning,sitzmann2019scene,Niemeyer_2020_CVPR}. These approaches assume opaque surfaces so that in each optimization iteration the gradient descent is with respect to the encoding of a single fragment per pixel. This is different from direct volume rendering applications, where the optimization needs to consider the contributions of many samples to a pixel color. This requires considering a large number of partial derivatives of pixel colors with respect to parameter or material changes, and to propagate them back into the volumetric field.   

The differentiable rendering framework Mitsuba 2~\cite{Mitsuba2} also provides a solution for direct volume rendering through Monte Carlo path tracing. However, Mitsuba directly applies so-called reverse-mode differentiation, which requires all intermediate derivatives to be saved for backpropagation. Thus, the memory required in direct volume rendering applications quickly exceeds the available system memory. This limits the approach to small volumetric grids and a small number of volume interactions that cannot faithfully optimize for small-scale structures.
A follow-up work~\cite{NimierDavid2020Radiative} addresses this issue but limits the differentiability to volume densities and colors.

\subsection{Contribution}

This work presents a general solution for DiffDVR: differentiable Direct Volume Rendering using the emission-absorption model without multiple scattering. This requires analyzing approaches for automatic differentiation (AD) with respect to the specific requirements in direct volume rendering (DVR). So-called forward-mode approaches are efficient if the number of parameters is low, yet they become computationally too expensive with an increasing number of parameters, i.e., 
when optimizing for per-voxel densities in a volumetric field.  
The so-called reverse mode or adjoint mode records the operations and intermediate results in a graph structure. This structure is then traversed in reverse order during the backward pass that propagates the changes to the sample locations. However, this requires storing $O(kn)$ intermediate results, where $n$ is the number of pixels and $k$ the number of sample locations, and reversing the order of operations.

We show that a-priori knowledge about the operations performed in DVR can be exploited to avoid recording the operations in reverse-mode AD. We propose a custom computation kernel that inverts the order of operations in turn and derives the gradients used by AD. We further present a method for recomputing intermediate results via an analytic inversion of the light accumulation along the view rays. By this, intermediate results do not need to be recorded and the memory consumption of reverse-mode AD becomes proportional to $O(n)$. 

As our second contribution, we discuss a number of use cases in which AD is applied in volume rendering applications (\autoref{fig:teaser}). These use cases demonstrate the automatic optimization of external parameters of the rendering process, i.e., the camera and the TF. Here the 3D density field is not changed, but the optimization searches for the external parameters that---when used to render this field---yield an optimal solution of a problem-specific loss function. In addition, we cover problems where the optimization is with respect to the densities. I.e., the field values are optimized so that an image-based loss function---after rendering the optimized field---yields an optimal solution. We consider inverse tomography by restricting the rendering process to an absorption-only model without a TF and optimize the densities using given images of the field. For this case, we compare our method against algebraic reconstruction techniques~\cite{vanAarle2015astra,vanAarle2016astra} and Mitsuba~2~\cite{Mitsuba2,NimierDavid2020Radiative}. Beyond that, and for the first time to our best knowledge, we show how to incorporate TFs and an emission-absorption model into tomographic reconstruction and deal with the resulting non-convex optimization problem.


DiffDVR is written in C++ and CUDA, and it provides seamless interoperability with PyTorch for a simple embedding into existing training environments with complex, potentially network-based loss functions. The code is made publicly available under a BSD license\footnote{\url{https://github.com/shamanDevel/DiffDVR}}.

\section{Related Work}

\paragraph*{Differentiable Rendering}
A number of differentiable renderers have been introduced for estimating scene parameters or even geometry from reference images, for example, under the assumption of local illumination and smooth shading variations ~\cite{loper2014opendr,petersen2019pix2vex,rhodin2015versatile,kato2018neural}, or via edge sampling to cope with discontinuities at visibility boundaries ~\cite{li2018differentiable}. Scattering parameters of homogeneous volumes have been estimated from observed scattering pattern ~\cite{gkioulekas2013inverse}.
Recently, Nimier-David~\etAl\ proposed Mitsuba 2~\cite{Mitsuba2}, a fully-differentiable physically-based Monte-Carlo renderer. Mitsuba 2 also handles volumetric objects, yet it requires storing intermediate results during the ray sampling process at each sampling point. This quickly exceeds the available memory and makes the approach unfeasible for direct volume rendering applications. Later, the authors have shown how to avoid storing the intermediate results~\cite{NimierDavid2020Radiative}, by restricting the parameters that can be derived to, e.g., only shading and emission.
However, these methods are tailored for path tracing with multiple scattering and rely on Monte-Carlo integration with delta tracking. This makes them prone to noise and leads to long computation times compared to classical DVR methods without scattering.  
Our method, in contrast, does not require storing intermediate results and can, thus, use large volumes with arbitrary many sampling steps without resorting to a restricted parameter set.
Furthermore, it does not impose restrictions on the parameters of the volume rendering process that can be differentiated.

\paragraph*{Parameter Optimization for Volume Visualization}
An interesting problem in volume visualization is the automatic optimization of visualization parameters like the viewpoint, the TF, or the sampling stepsize that is required to convey the relevant information in the most efficient way. This requires at first hand an image-based loss function that can be used to steer the optimizer toward an optimal parameter setting.  
To measure a viewpoint's quality from a rendered image, loss functions based on image entropy~\cite{ji2006dynamic,vazquez2008representative,tao2009structure,chen2010information} or image similarity~\cite{tao2016similarity,yang2019deep} have been used. For volume visualization, the relationships between image entropy and voxel significance ~\cite{bordoloi2005view} as well as importance measures of specific features like isosurfaces~\cite{takahashi2005feature} have been considered. None of these methods, however, considers the rendering process in the optimization process. Instead, views are first generated from many viewpoints, e.g., by sampling via the Fibonacci sphere algorithm~\cite{marques2013fibonacci}, and then the best view regarding the used loss function is determined. We envision that by considering the volume rendering process in the optimization, more accurate and faster reconstructions can be achieved.

Another challenging problem is the automatic selection of a ``meaningful'' TF for a given dataset, as the features to be displayed depend highly on the user expectation.
Early works attempted to find a good TF using clusters in histograms of statistical image properties~\cite{haidacher2010volume} or fitting visibility histograms~\cite{correa2010visibility}.
Others have focused on guiding an explorative user interaction ~\cite{zhou2013transfer,maciejewski2012abstracting}, also by using neural networks~\cite{berger2018generative}.
For optimizing a TF based on information measures, Ruiz~\etAl~\cite{ruiz2011automatic} proposed to bin voxels of similar density and match their visibility distribution from multiple viewpoints with a target distribution defined by local volume features. 
For optimization, the authors employ a gradient-based method where the visibility derivatives for each density bin are approximated via local linearization.

Concerning the performance of direct volume rendering, it is crucial to determine the minimum number of data samples that are required to accurately represent the volume. 
In prior works, strategies for optimal sampling in screen-space have been proposed, for instance, based on perceptual models~\cite{Bolin98-perceptuallybasedsampling}, image saliency maps~\cite{painter1989antialiased}, entropy-based measures~\cite{xu2005adaptive}, temporal history~\cite{martschinke2019AdaptivePathTracing}, or using neural networks~\cite{weiss2019isosuperres,weiss2020adaptivesampling}.
Other approaches adaptively change the sampling stepsize along the view rays to reduce the number of samples in regions that do not contribute much to the image \cite{Danskin92-fastvolren,kratz2011adaptive, Campagnolo2015-rayadapt,morrical2019spaceskipping}.
DiffDVR's capability to compute gradients with respect to the stepsize gives rise to a gradient-based adaptation using image-based loss functions instead of gradient-free optimizations or heuristics. 

\paragraph*{Neural Rendering}
As an alternative to classical rendering techniques that are adapted to make them differentiable, several works have proposed to replace the whole rendering process with a neural network.
For a general overview of neural rendering approaches let us refer to the recent summary article by Tewari~\etAl~\cite{tewari2020state}.
For example, RenderNet proposed by Nguyen-Phuoc~\etAl~\cite{nguyen2018rendernet} replaces the mesh rasterizer with a combination of convolutional and fully connected networks.
In visualization, the works by Berger~\etAl~\cite{berger2018generative} and He~\etAl~\cite{he2019insitunet} fall into the same line of research. The former trained a network on rendered images and parameters of the rendering process, and use the network to predict new renderings by using only the camera and TF parameters. The latter let a network learn the relationships between the input parameters of a simulation and the rendered output, and then used this network to skip the rendering process and create images just from given input parameters. 

\section{Background}\label{sec:pre}
In the following, we review the fundamentals underlying DVR using an optical emission-absorption model~\cite{max1995optical}. Then we briefly summarise the foundation of Automatic Differentiation (AD), a method to systematically compute derivatives for arbitrary code~\cite{BARTHOLOMEWBIGGS2000171}.
\subsection{Direct Volume Rendering Integral}\label{sec:pre:dvr}

Let $V : \R^3 \rightarrow [0,1]$ be the scalar volume of densities and let $r : \R^+ \rightarrow \R^3$ be an arc-length parameterized ray through the volume.
Let $\tau: [0,1]\rightarrow \R_0^+$ be the absorption and $C: [0,1]\rightarrow \R_0^+$ the self-emission due to a given density.
Then, the light intensity reaching the eye is given by 
\begin{equation}
    L(a,b) = \int_a^b g(V(r(t))) e^{-\int_a^t\tau(V(r(u)))\dd u} \dd t,
    \label{eq:integral2}
\end{equation}
were the exponential term is the transparency of the line segment from $t=a$, the eye, to $b$, the far plane, and $g(v) = \tau(v)C(v)$ is the emission. The transparency is one if the medium between $a$ and $b$ does not absorb any light and approaches zero for complete absorption.

We assume that the density volume is given at the vertices $\mathbf{v}_i$ of a rectangular grid, 
and the density values are obtained via 
\emph{trilinear} interpolation.
The functions $\tau$ and $C$ define the mapping from density to absorption and emission. 
We assume that both functions are discretized into $R$ regularly spaced control points with linear interpolation in between. This is realized on the GPU as a 1D texture map $T$ with hardware-supported linear interpolation.

For arbitrary mappings of the density to absorption and emission, the volume rendering integral in \autoref{eq:integral2} cannot be solved analytically. Instead, it is approximated by discretizing the ray into $N$ segments over which the absorption $\alpha_i$ and emission $L_i$ are assumed constant. 
We make use of the Beer-Lambert model $\alpha_i=1-\exp(-\Delta t \tau(d_i))$, where $d_i$ is the sampled volume density, to approximate a segment's transparency.
This leads to a Riemann sum which can be computed in front-to-back order using iterative application of alpha-blending, i.e., $L = L + (1-\alpha) L_i$, and $\alpha = \alpha+(1-\alpha)\alpha_i$.


\subsection{Automatic Differentiation}\label{sec:pre:wengert}
The evaluation of any program for a fixed input can be expressed as a computation graph, a directed acyclic graph where the nodes are the operations and the edges the intermediate values. Such a computation graph can be reformulated as a linear sequence of operations, also called a Wengert list~\cite{wengert1964simple,BARTHOLOMEWBIGGS2000171},
\begin{equation}
\begin{aligned}
    \bm{x}_0 &= \text{const} \\
    \bm{x}_1 &= f_1(\bm{x}_0, \bm{w}_1) \\
    \bm{x}_2 &= f_2(\bm{x}_1, \bm{w}_2) \\[-1mm]
    & \hdots\\[-1mm]
    \bm{x}_{\text{out}} &= f_k(\bm{x}_{k-1}, \bm{w}_k)
\end{aligned}
\label{eq:program}
\end{equation}
where the $\bm{w}_i$'s $\in\R^{p}$ are the external parameters of the operations of size $p$ and the $\bm{x}_i$'s $\in\R^{n}$ refer to the state of intermediate results after the i-th operation of size $n$.
The output $\bm{x}_{\text{out}}\in\R^m$ has size $m$. Note here that in DiffDVR, $n$ and $k$ are usually large, i.e., $n$ is in the order of the number of pixels and $k$ in the order of the number of sampling points along the view rays. The output $\bm{x}_{\text{out}}$ is a scalar ($m=1$), computed, for example, as the average per-pixel loss over the image.
The goal is then to compute the derivatives 
    $\frac{\dd \bm{x}_\text{out}}{\dd \bm{w}_i}$.
The basic idea is to split these derivatives into simpler terms for each operation using the chain rule. For example, assuming univariate functions and $w_1$ the only parameter of interest, the chain rule yields
\begin{equation}
    x_3 = f_3(f_2(f_1(w_1))) \ \Rightarrow \ x_3'=f_3'(f_2(f_1(x)))\, f_2'(f_1(x))\, f_1'(x).
    \label{eq:chainrule}
\end{equation}


There are two fundamentally different approaches to automatically evaluate the chain rule, which depend on the order of evaluations. If the product in the above example is evaluated left-to-right, the derivatives are propagated from bottom to top in \autoref{eq:program}. This gives rise to the adjoint- or backward-mode differentiation (see \autoref{sec:pre:autodiff:backward}). If the product is evaluated right-to-left, the derivatives ``ripple downward'' from top to bottom in \autoref{eq:program}. This corresponds to the so-called forward-mode differentiation (see \autoref{sec:pre:autodiff:forward}). 


\section{AD for Direct Volume Rendering}\label{AD_DVR}

Now we introduce the principal procedure when using AD for DiffDVR and hint at the task-dependent differences when applied for viewpoint optimization (\autoref{sec:app:viewpoint}), TF reconstruction (\autoref{sec:app:tf}) and volume reconstruction (\autoref{sec:app:volume} and \autoref{sec:app:color}). We further discuss computational aspects and memory requirements of AD in volume rendering applications and introduce the specific modifications to make DiffDVR feasible.

\subsection{The Direct Volume Rendering Algorithm}\label{sec:autodiff:concrete}

In direct volume rendering, the pixel color represents the accumulated attenuated emissions at the  sampling points along the view rays. In the model of the Wengert list (see \autoref{eq:program}), a function $f_i$ is computed for each sample. Hence, the number of operations $k$ is proportional to the overall number of samples along the rays.
The intermediate results $\bm{x}_i$ are rgb$\alpha$ images of the rendered object up to the $i$-th sample, i.e., $\bm{x}_i$ is of size $n=\text{W}*\text{H}*4$, where $W$ and $H$, respectively, are the width and height of the screen.
The last operation $f_k$ in the optimization process is the evaluation of a scalar-valued loss function. Thus, the size of the output variable is $m=1$.
The parameters $\bm{w}_i$ depend on the use case. For instance, in viewpoint optimization, the optimization is for the longitude and latitude of the camera position, i.e., $p=2$.
When reconstructing a TF, the optimization is for the $R$ rgb$\alpha$ entries of the TF, 
i.e., $p=4R$.

The DVR algorithm with interpolation, TF mapping, and front-to-back blending is shown in \autoref{alg:ftb}. 
For clarity, the variables in the algorithm are named by their function, instead of using $\bm{w}_i$ and $\bm{x}_i$ as in the Wengert list (\autoref{eq:program}).
In the Wengert list model, the step size $\Delta t$, the camera intrinsics $cam$, the TF $T$, and the volume density $V$ are the parameters $\bm{w}_i$. The other intermediate variables are represented by the states $\bm{x}_i$.
Each function operates on a single ray but is executed in parallel over all pixels. 
\begin{algorithm}[H]
\caption{Direct Volume Rendering Algorithm}\label{alg:ftb}
\begin{algorithmic}[1]
    \Parameters stepsize $\Delta t$, camera $cam$, TF $T$, volume $V$
    \Input $u\!v$ the pixel positions where to shoot the rays
    \State $color_{i}=0$ \Comment{initial foreground color}
    \State $x_o,\omega = f_{\text{camera}}(u\!v, cam)$ \Comment{start $x_o$ and direction $\omega$ for all rays}
    \For{$i=0, ..., N-1$}
        \State $x_{i} = x_o + i \Delta t \omega$ \Comment{current position along the ray}
        \State $d_{i} = f_{\text{interpolate}}(x_{i}, V)$ \Comment{Trilinear interpolation}
        \State $c_{i} = f_{\text{TF}}(d_{i}, T)$ \Comment{TF evaluation}
        \State $color_{i+1} = f_{\text{blend}}(color_{i}, c_{i})$ \Comment{blending of the sample}\label{alg:ftb:blending}
    \EndFor
    \State $\bm{x}_{\text{out}} = f_{\text{loss}}(color_{N})$ \Comment{Loss function on the output rgb$\alpha$ image}
\end{algorithmic}
\end{algorithm}
When \autoref{alg:ftb} is executed, the operations form the computational graph. 
AD considers this graph to compute the derivatives of $\bm{x}_{\text{out}}$ with respect to the parameters $\Delta t, cam, T, \text{ and } V$, so that the changes that should be applied to the parameters to optimize the loss function can be computed automatically.
Our implementation allows for computing derivatives with respect to all parameters, yet due to space limitations, we restrict the discussion to the computation of derivatives of $\bm{x}_{\text{out}}$ with respect to the camera $cam$, the TF $T$ and the volume densities $V$.
In the following, we discuss the concrete implementations of forward and adjoint differentiation to compute these derivatives. 

\subsection{Forward Differentiation}\label{sec:pre:autodiff:forward}
On the elementary level, the functions in \autoref{alg:ftb} can be expressed as a sequence of scalar arithmetic operations like $c=f(a,b)=a*b$.
In forward-mode differentiation~\cite{neidinger2010introduction,BARTHOLOMEWBIGGS2000171},
every variable is replaced by the associated \emph{forward variable}
\begin{equation}
    \tilde{a} = \left\langle a, \frac{\dd a}{\dd w} \right\rangle , \ 
    \tilde{b} = \left\langle b, \frac{\dd b}{\dd w} \right\rangle ,
\end{equation}
i.e., tuples of the original value and the derivative with respect to the parameter $w$ that is optimized.
Each function $c=f(a,b)$ is replaced by the respective forward function
\begin{equation}
    \tilde{c} = \tilde{f}(\tilde{a}, \tilde{b}) = \left\langle f(a,b), \frac{\partial f}{\partial a}\frac{\dd a}{\dd w} + \frac{\partial f}{\partial b}\frac{\dd b}{\dd w} \right\rangle .
\end{equation}
Constant variables are initialized with zero, $\tilde{x}_\text{const}=\langle x_\text{const}, 0\rangle$, and parameters for which to trace the derivatives are initialized with one, $\tilde{w}=\langle w, 1\rangle$.
If derivatives for multiple parameters should be computed, the tuple of forward variables is extended.

Forward differentiation uses a custom templated datatype for the forward variable and operator overloading. Each variable 
is wrapped in an instance of this datatype, called \mintinline{c++}|fvar|,
which stores the derivatives with respect to up to $p$ parameters along with their current values.
\begin{mdframed}[backgroundcolor=gray!5!white,roundcorner=5pt]{\small
\begin{minted}{c++}
template<typename T, int p>
struct fvar
{
    T value;
    T derivatives[p];
};
\end{minted}
}\end{mdframed}
\noindent 
Next, operator overloads are provided for all common arithmetic operations and their gradients. 
For example, multiplication is implemented similar to:
\begin{mdframed}[backgroundcolor=gray!5!white,roundcorner=5pt]{\small
\begin{minted}{c++}
template<typename T, int p>
fvar<T, p> operator*(fvar<T, p> a, fvar<T, p> b) 
{
    fvar<T, P> c; //to store c = a*b and derivatives
    c.value = a.value * b.value;
    for (int i=0; i<p; ++i) { //partial derivatives
        c.derivative[i] = a.value*b.derivative[i] 
            + b.value*a.derivative[i];
    }
    return c;
}
\end{minted}
}\end{mdframed}
The user has to write the functions in such a way that arbitrary input types are possible, i.e., regular floats or instances of \mintinline{c++}|fvar|, via C++ templates. All intermediate variables are declared with type \mintinline{c++}|auto|. This allows the compiler to use normal arithmetic if no gradients are propagated, but when forward variables with gradients are passed as input, the corresponding operator overloads are chosen.

\begin{figure}[t]
    \centering
    \includegraphics[width=\linewidth,trim=0 90 170 30,clip]{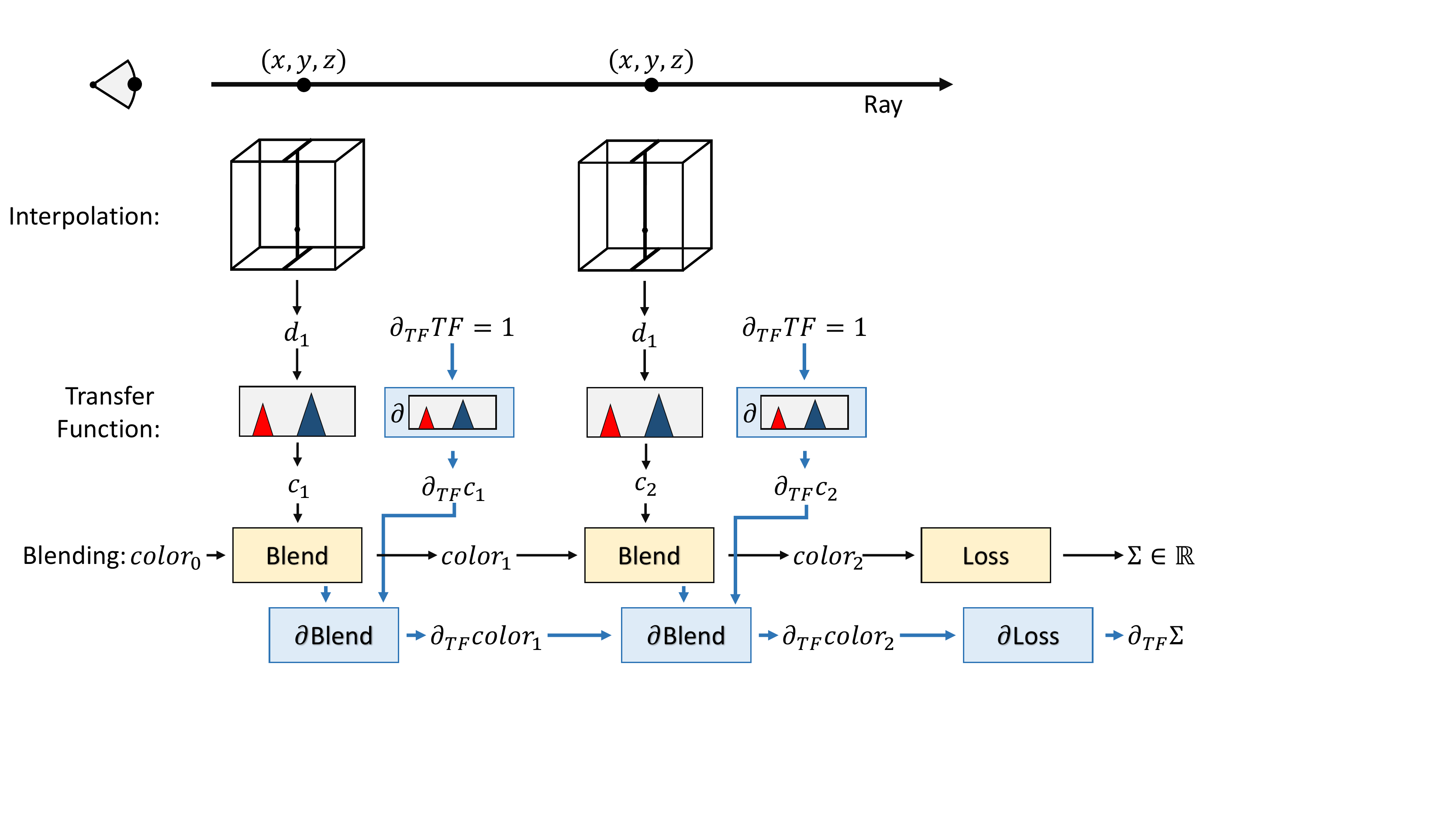}\vspace*{-2mm}
    \caption{Schematic representation of the forward method for TF reconstruction. Gradients are stored in the forward variables (blue), and parameter values are propagated simultaneously.}
    \label{fig:forward}
\end{figure}

As an example (see \autoref{fig:forward} for a schematics), let us assume that derivatives should be computed with respect to a single entry in a 1D texture-based TF, e.g., the red channel of the first texel $T_{0,\text{red}}$.
When loading the TF from memory, $T_{0,\text{red}}$ is replaced by $\tilde{T}_{0,\text{red}} = \langle T_{0,\text{red}}, 1 \rangle$, i.e., it is wrapped in an instance of \mintinline{c++}|fvar| with the derivative for that parameter set to $1$.
\autoref{alg:ftb} executes in the normal way until 
$T_{0,\text{red}}$ is encountered in the code for the TF lookup. Now, the operator overloading mechanism selects the forward function instead of the normal non-differentiated function.
The result is not a regular color $c_i$, but the forward variable of the color $\tilde{c}_i$. All following functions (i.e., the blend and loss function) continue to propagate the derivatives.
In contrast, if derivatives should be computed with respect to the camera, already the first operation requires tracing the derivatives with \mintinline{c++}|fvar|.

It is worth noting that in the above example only the derivative of one single texel in the TF is computed. This process needs to be repeated for each texel, respectively each color component of each texel, by extending the array \mintinline{c++}|fvar::derivatives| to store the required number of $p$ parameters. Notably, for input data that is high dimensional, like TFs or a 3D volumetric field, forward differentiation becomes unfeasible. For viewpoint selection, on the other hand, where only two parameters are optimized, forward differentiation can be performed efficiently.

The computational complexity of the forward method scales linearly with the number of parameters $p$, as they have to be propagated through every operation. However, as every forward variable directly stores the derivative of that variable w.r.t. the parameters, gradients for an arbitrary number of outputs $m$ can be directly realized. Furthermore, the memory requirement is proportional to $O(np)$, as only the current state needs to be stored.

\begin{figure}[b]
    \centering
    \includegraphics[width=\linewidth,trim=0 65 170 30,clip]{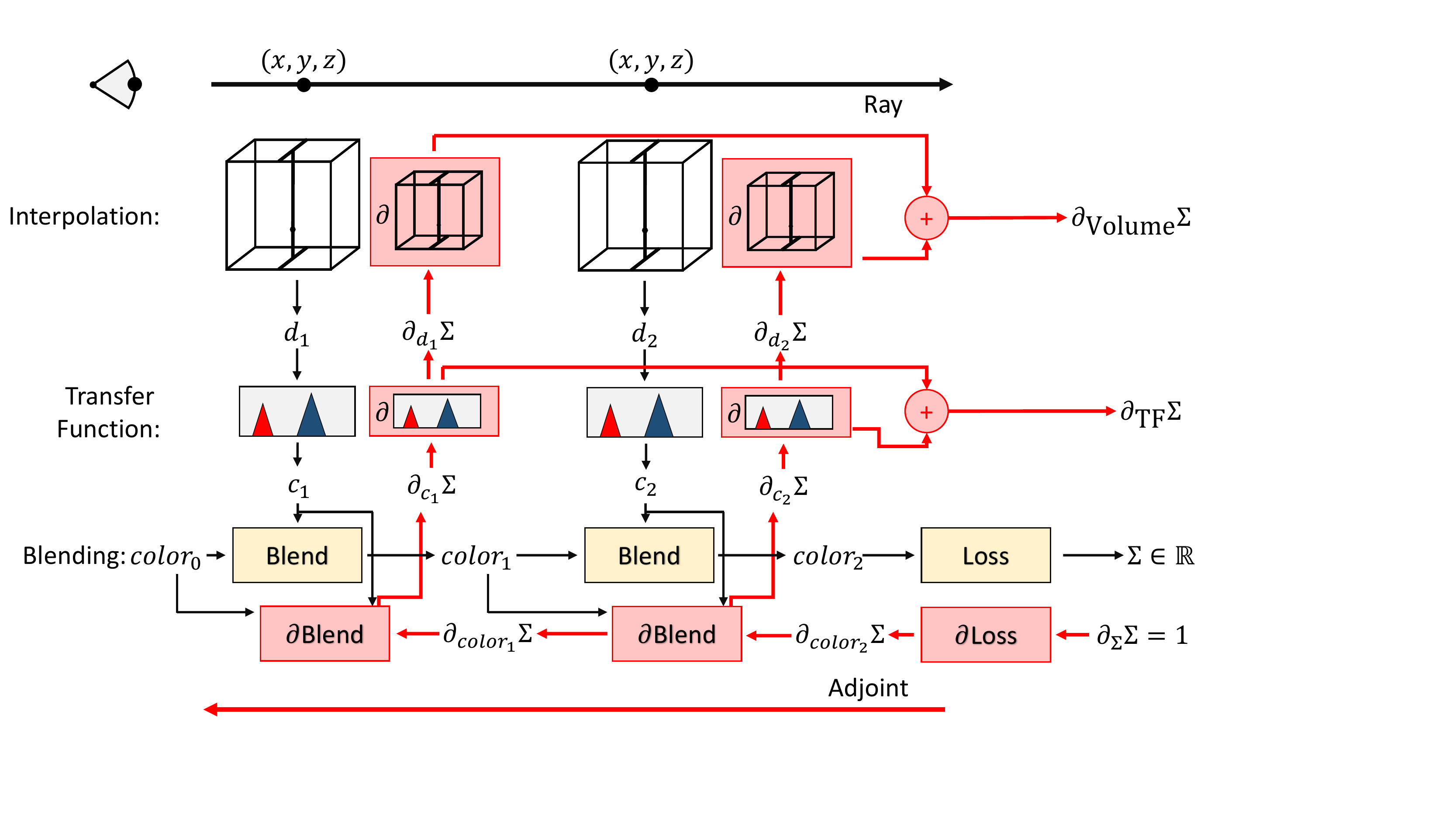}\vspace*{-2mm}%
    \caption{Schematic representation of the adjoint method for density and TF reconstruction. Gradients in the adjoint variables (red) are propagated backward through the algorithm. A circled $+$ indicates the summation of the gradients over all steps and rays.}
    \label{fig:adjoint}
\end{figure}

\subsection{Adjoint Differentiation}\label{sec:pre:autodiff:backward}
Adjoint differentiation~\cite{McNamara.2004}, also called the adjoint method, backward or reverse mode differentiation, or backpropagation, evaluates the chain rule in the inverse order than forward differentiation.
%
For each variable $\bm{x}_i$, the associated \emph{adjoint variable}
\begin{equation}
    \hat{\bm{x}}_i=\frac{\partial x_{\text{out}}}{\partial \bm{x}_i} \ , \
    \hat{\bm{w}}_i=\frac{\partial x_{\text{out}}}{\partial \bm{w}_i},
    \label{eq:adjoin1}
\end{equation}
stores the derivative of the final output with respect to the current variable.
Tracing the derivatives starts by setting $\hat{x}_\text{out}=1$. Then, the adjoint variables are tracked backward through the algorithm, called the \emph{backward pass}. This is equivalent to evaluating the chain rule \autoref{eq:chainrule} from left to right, instead of right to left as in the forward method. Let $c=f(a,b)$ be again our model function, then the adjoint variables $\hat{a},\hat{b}$ are computed from $\hat{c}$ as
\begin{equation}
    \hat{a} = \left(\frac{\partial f}{\partial a}\right)^T \hat{c} , \ \hat{b} = \left(\frac{\partial f}{\partial b}\right)^T \hat{c} .
    \label{eq:simpleadjoint}
\end{equation}
This process is repeated from the last operation to the first operation, giving rise to the \emph{adjoint code}.
At the end, one arrives again at the derivatives with respect to the parameters $\hat{\bm{w}}=\frac{\partial x_{\text{out}}}{\partial \bm{w}}$. 
If a parameter is used multiple times, either along the ray or over multiple rays, the adjoint variables are summed up. The reverted evaluation of the DVR algorithm with the gradient propagation from \autoref{eq:simpleadjoint} is sketched in \autoref{alg:adj-ftb}.
A schematic visualization is shown in \autoref{fig:adjoint}.

Because the adjoint method requires reversing the order of operation, simple operator overloading as in the forward method is no longer applicable.
Common implementations of the adjoint method like TensorFlow~\cite{abadi2016tensorflow} or PyTorch~\cite{NEURIPS2019_9015} record the operations in a computation graph, which is then traversed backward in the backward pass.
As it is too costly to record every single arithmetic operation, high-level functions like the evaluation of a single layer in neural networks are treated as atomic, and only these are recorded. Within such a high-level function, the order of operations is known and the adjoint code using \autoref{eq:simpleadjoint} is manually derived and implemented.
We follow the same idea and treat the rendering algorithm as one unit and manually derive the adjoint code.

\begin{algorithm}[t]
\caption{Adjoint Code of the DVR Algorithm. Each line corresponds to a line in ~\autoref{alg:ftb} in reverse order.}\label{alg:adj-ftb}
\begin{algorithmic}[1]
    \Parameters stepsize $\Delta t$, camera $cam$, TF $T$, volume $V$
    \Input the adjoint of the output $\hat{\bm{x}}_{\text{out}}$
    \Statex all intermediate adjoint variables are initialized with $0$
    \State $\hat{color}_{N} \pluseq \partial f_{\text{loss}}(color_{N})^T \hat{\bm{x}}_{\text{out}}$
    \For{$i=N-1, ..., 0$}
        \State $\hat{color}_{i}, \hat{c}_{i} \pluseq \partial f_{\text{blend}}(color_{i}, c_{i})^T \hat{color}_{N}$
        \State $\hat{d}_{i}, \hat{T} \pluseq \partial f_{\text{TF}}(d_{i}, T)^T \hat{c}_{i}$
        \State $\hat{x}_i, \hat{V} \pluseq \partial f_{\text{interpolate}}(x_i, V)^T \hat{d}_{i}$
        \State $\hat{x}_o \pluseq \hat{x}_i \ , \ \hat{\Delta t} \pluseq i\omega^T \hat{x}_i\ , \ \hat{\omega} \pluseq i\Delta t\hat{x}_i$
    \EndFor
    \State $\hat{cam} \pluseq \partial f_{\text{camera}}(u\!v, cam)^T [x_o;  \omega]$
    \State $\hat{color}_0$ is ignored
    \Output $\hat{\Delta t}, \hat{cam}, \hat{T}, \hat{V}$
\end{algorithmic}
\end{algorithm}


\subsection{The Inversion Trick}

One of the major limitations of the adjoint method is its memory consumption because the input values for the gradient computations need to be stored. 
For example, the blending operation (line \ref{alg:ftb:blending} in \autoref{alg:ftb}) is defined as follows: Let $\alpha, C$ be the opacity and rgb-emission at the current sample, i.e., the components of $c_i$, and let $\alpha^{(i)}, C^{(i)}$ be the accumulated opacity and emission up to the current sample, i.e., the components of $color_{i}$ in \autoref{alg:ftb}.
Then, the next opacity and emission is given by front-to-back blending
\begin{equation}
    \begin{aligned}
    C^{(i+1)} &= C^{(i)}+(1-\alpha^{(i)})C \\
    \alpha^{(i+1)} &= \alpha^{(i)}+(1-\alpha^{(i)})\alpha .
    \end{aligned}
    \label{eq:blend-forward}
\end{equation}
In the following adjoint code with $\hat{\alpha}^{(i+1)}, \hat{C}^{(i+1)}$ as input it can be seen that the derivatives again require the input values. 
\begin{equation}
    \begin{aligned}
    \hat{\alpha} &= (1-\alpha^{(i)})\hat{\alpha}^{(i+1)} , \ \hat{C}=(C-\alpha^{(i)})\hat{C}^{(i+1)} ,\\
    \hat{\alpha}^{(i)} &= (1-\alpha)\hat{\alpha}^{(i+1)} - C\cdot\hat{C}^{(i+1)} , \\
    \hat{C}^{(i)} &= \hat{C}^{(i+1)} .
    \end{aligned}
    \label{eq:adj-blend-forward}
\end{equation}
Therefore, the algorithm is first executed in its non-adjoint form, and the intermediate colors are stored with the computation graph. This is called the \emph{forward pass}.
During the backward pass, when the order of operations is reversed and the derivatives are propagated (the adjoint code), the intermediate values are reused.
In DVR, intermediate values need to be stored at every step through the volume. Thus, the memory requirement scales linearly with the number of steps and quickly exceeds the available memory.
To overcome this limitation, we propose a method that avoids storing the intermediate colors after each step and, thus, has a constant memory requirement.  

\begin{figure}
    \centering
    \subfloat[No Inversion]{
        \includegraphics[width=0.48\linewidth,valign=b]{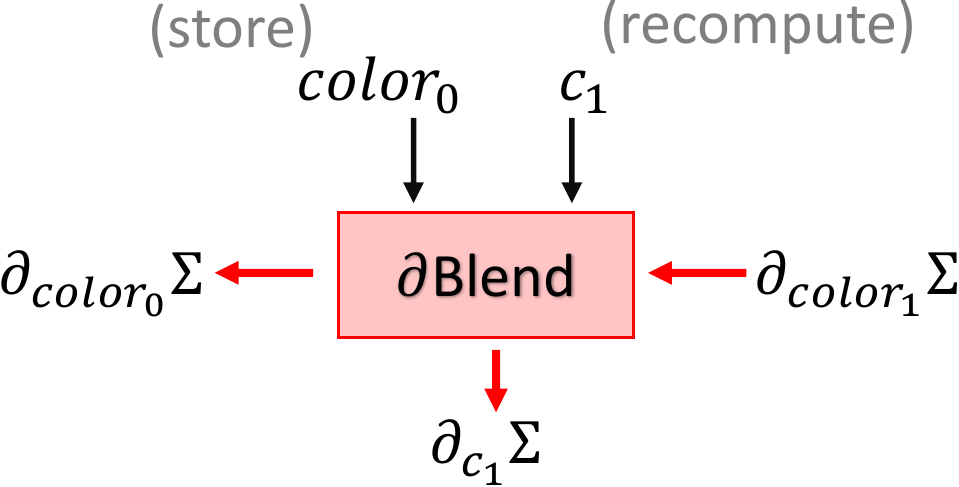}
    }~~~
    \subfloat[With Inversion]{
        \includegraphics[width=0.48\linewidth,valign=b]{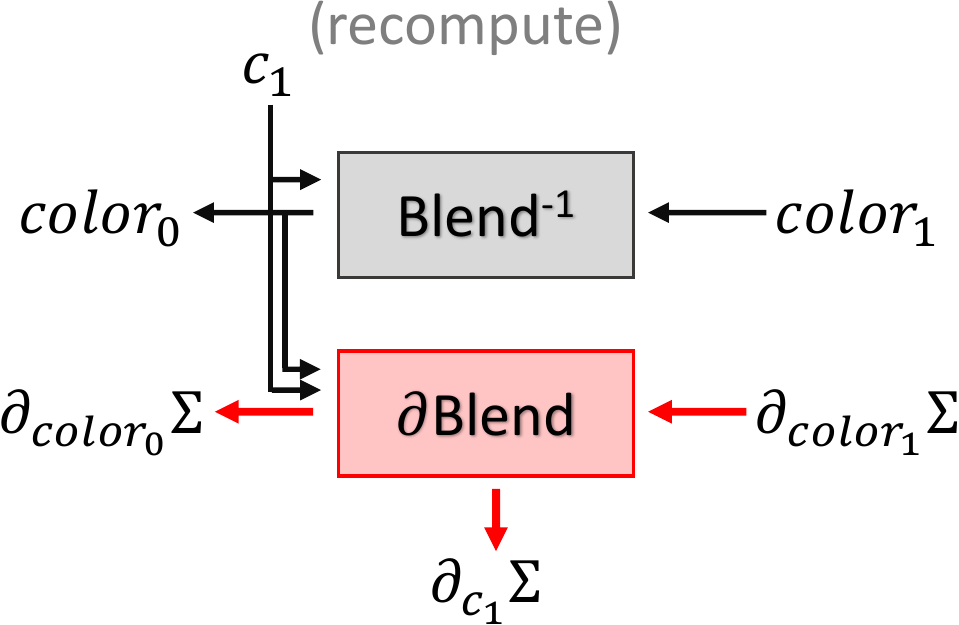}
    }%
    \caption{(a) To compute the current contribution $c_i$, intermediate accumulated colors $\text{color}_i$ need to be stored for every step along the ray. (b) The inversion trick enables to reconstruct $\text{color}_i$ from $\text{color}_{i+1}$. Thus, only the final color used in the loss function needs to be stored.}
    \label{fig:adjoint:inversion}
\end{figure}

We exploit that the blending step is invertible (see \autoref{fig:adjoint:inversion}): If $\alpha^{(i+1)}, C^{(i+1)}$ are given and the current sample is recomputed to obtain $\alpha$ and $C$, $\alpha^{(i)}, C^{(i)}$ can be reconstructed as
\begin{equation}
    \begin{aligned}
    \alpha^{(i)} &= \frac{\alpha - \alpha^{(i+1)}}{\alpha - 1} \\
    C^{(i)} &= C^{(i+1)} - (1-\alpha^{(i)})C .
    \end{aligned}
    \label{eq:inv-blend-forward}
\end{equation}
With \autoref{eq:inv-blend-forward} and $\alpha < 1$, the adjoint pass can be computed with constant memory by re-evaluating the current sample $c_i$ and reconstructing $color_i$ instead of storing the intermediate results. 
Thus, only the output color used in the loss function needs to be stored, while all intermediate values are recomputed on-the-fly. Note that $\alpha = 1$ is not possible in practice, since it requires the absorption stored in the TF to be at infinity. 

\begin{figure*}[t]
    \centering
    \subfloat[]{
        \includegraphics[width=0.3428\textwidth]{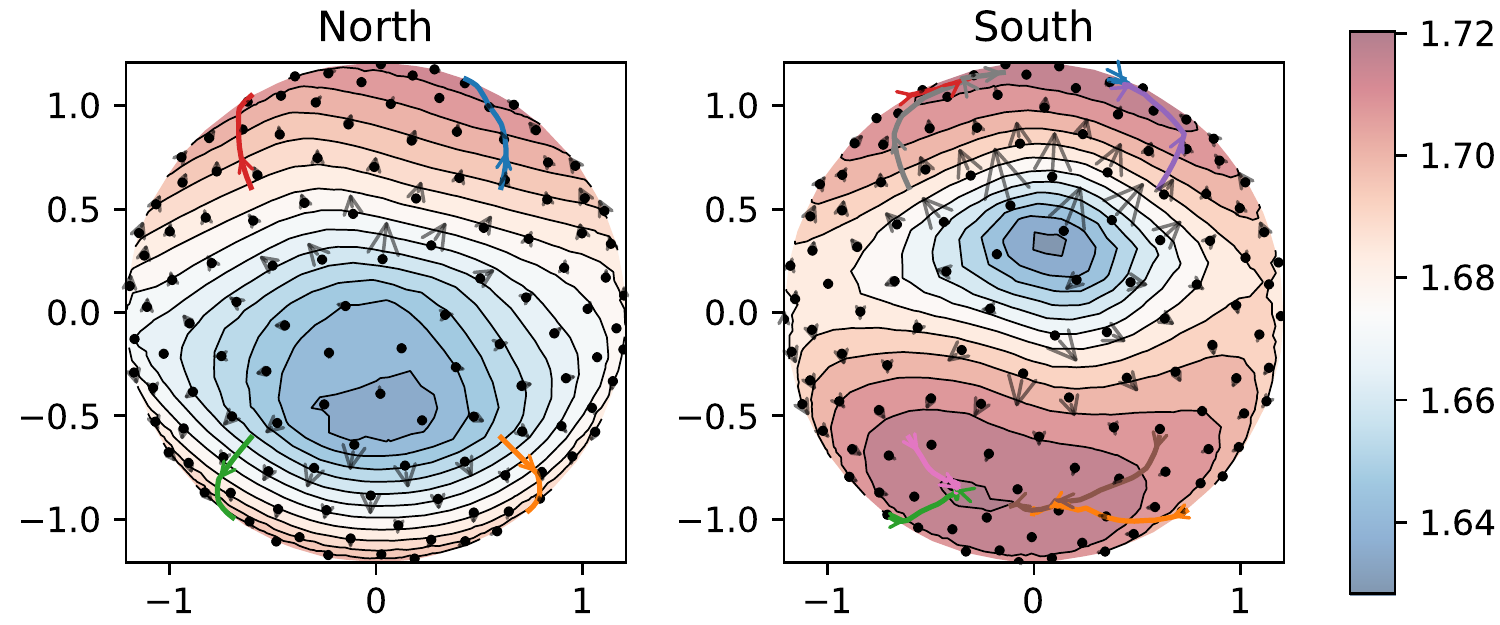}
    }~
    \subfloat[]{
        \includegraphics[width=0.3428\textwidth,valign=b]{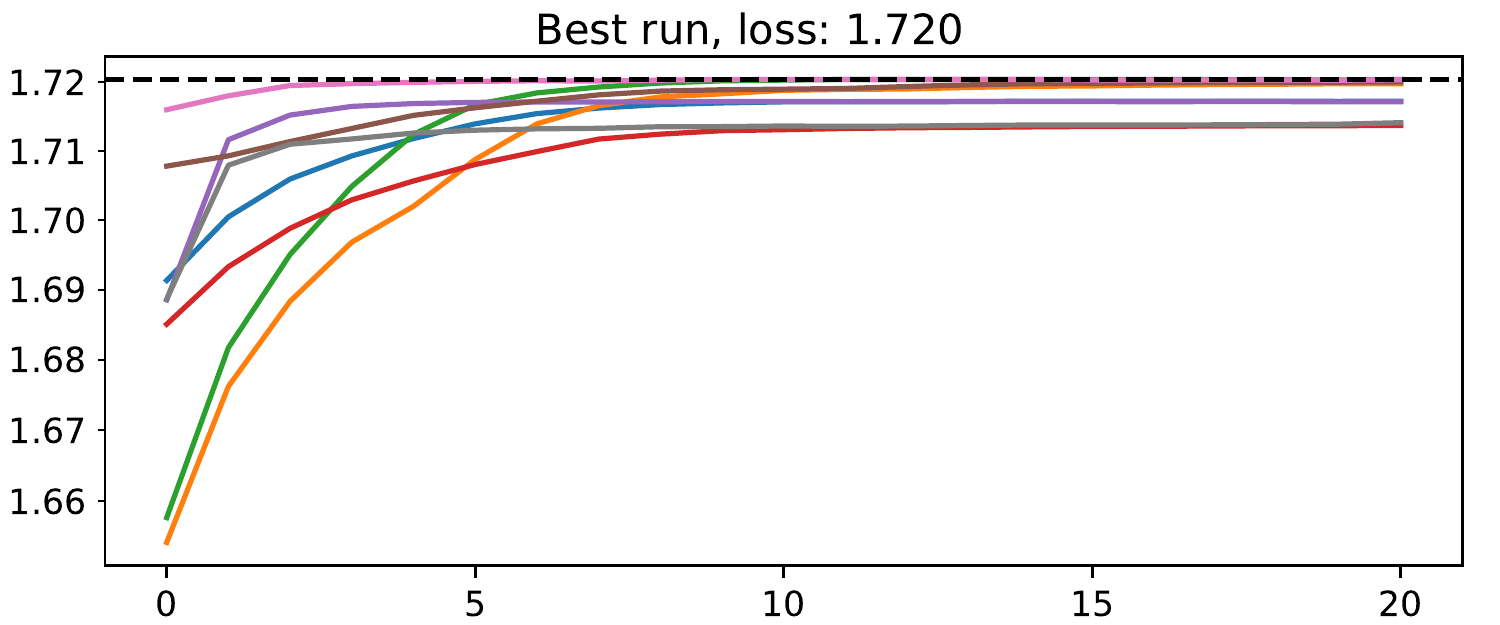}
    }~
    \subfloat[]{
        \includegraphics[width=0.28571\textwidth,valign=b]{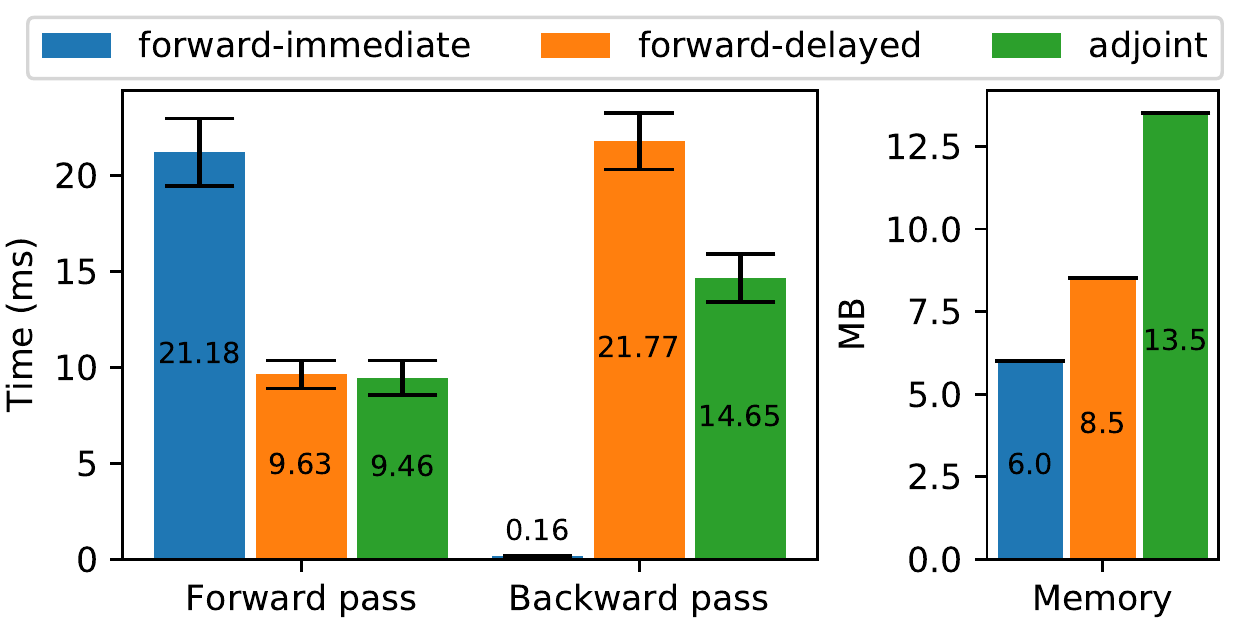}
    }
    \vspace*{-1em}
    \caption{Best viewpoint selection using maximization of visual entropy. The tooth dataset (\autoref{fig:teaser}a) is rendered from different viewpoints on a surrounding sphere. (a) Color coding of loss values for viewpoints on the northern and southern hemispheres, with isocontours (black lines) of the loss and local gradients with respect to the longitude and latitude of the camera position at uniformly sampled positions (black dots with arrows). Eight optimization runs (colored paths on the surface) are started at uniformly seeded positions and optimized in parallel. (b) The runs converge to three clusters of local minima. The cluster with the highest entropy ($1.72$) coincides with the best value from $256$ sampled entropies. For the best run, the start view, as well as some intermediate views and the final result, are shown in \autoref{fig:teaser}a. (c) Timings and memory consumption show that forward differences approximately double the runtime, but are faster and require less memory than the adjoint method.}
    \label{fig:app:camera:tooth2d}
\end{figure*}

In the implementation, and indicated by the circled $+$ in \autoref{fig:adjoint}, the adjoint variables for the parameters are first accumulated per ray into local registers (camera, stepsize, volume densities) or shared memory (TF). Then, the variables are accumulated over all rays using global atomic functions. This happens once all rays have been traversed (camera, stepsize, transfer function) or on exit of the current cell (volume densities).

Because the adjoint variables carry only the derivatives of the output, but not of the parameters, the computational complexity is largely constant in the number of parameters. For example, in TF optimization (\autoref{sec:app:tf}) only the derivative of the currently accessed texel is computed when accessed in the adjoint code of TF sampling. This is significantly different from the forward method, where the derivatives of all TF entries need to be propagated in every step. On the other hand, the adjoint method considers only a single scalar output in each backward pass, requiring multiple passes to support multi-component outputs. 
This analysis and the following example applications show that the forward method is preferable when optimizing for a low number of parameters like the camera position, while for applications such as TF optimization, which require the optimization of many parameters, the adjoint method has clear performance advantages.

DiffDVR is implemented as a custom CUDA operation in PyTorch~\cite{NEURIPS2019_9015}.
The various components of the DVR algorithm, like the parameter to differentiate or the type of TF, are selected via C++ template parameters. This eliminates runtime conditionals in the computation kernel.
To avoid pre-compiling all possible combinations, the requested configuration is compiled on demand via CUDA's JIT-compiler NVRTC~\cite{cuda-nvrtc} and cached between runs. This differs from, e.g., the Enoki library~\cite{Enoki} used by the Mitsuba renderer~\cite{Mitsuba2}, which directly generates Parallel Thread Code (PTX) for translation into GPU binary code.

\section{Applications}\label{sec:app}
In the following, we apply both AD modes for best viewpoint selection, TF reconstruction, and volume reconstruction. The results are analyzed both qualitatively and quantitatively. Timings are performed on a system running Windows 10 and CUDA 11.1 with an Intel Xeon 8x@3.60Ghz CPU, 64GB RAM, and an NVIDIA RTX 2070.

\subsection{Best Viewpoint Selection}\label{sec:app:viewpoint}


\begin{figure*}[t]
    \centering
    \includegraphics[width=\textwidth]{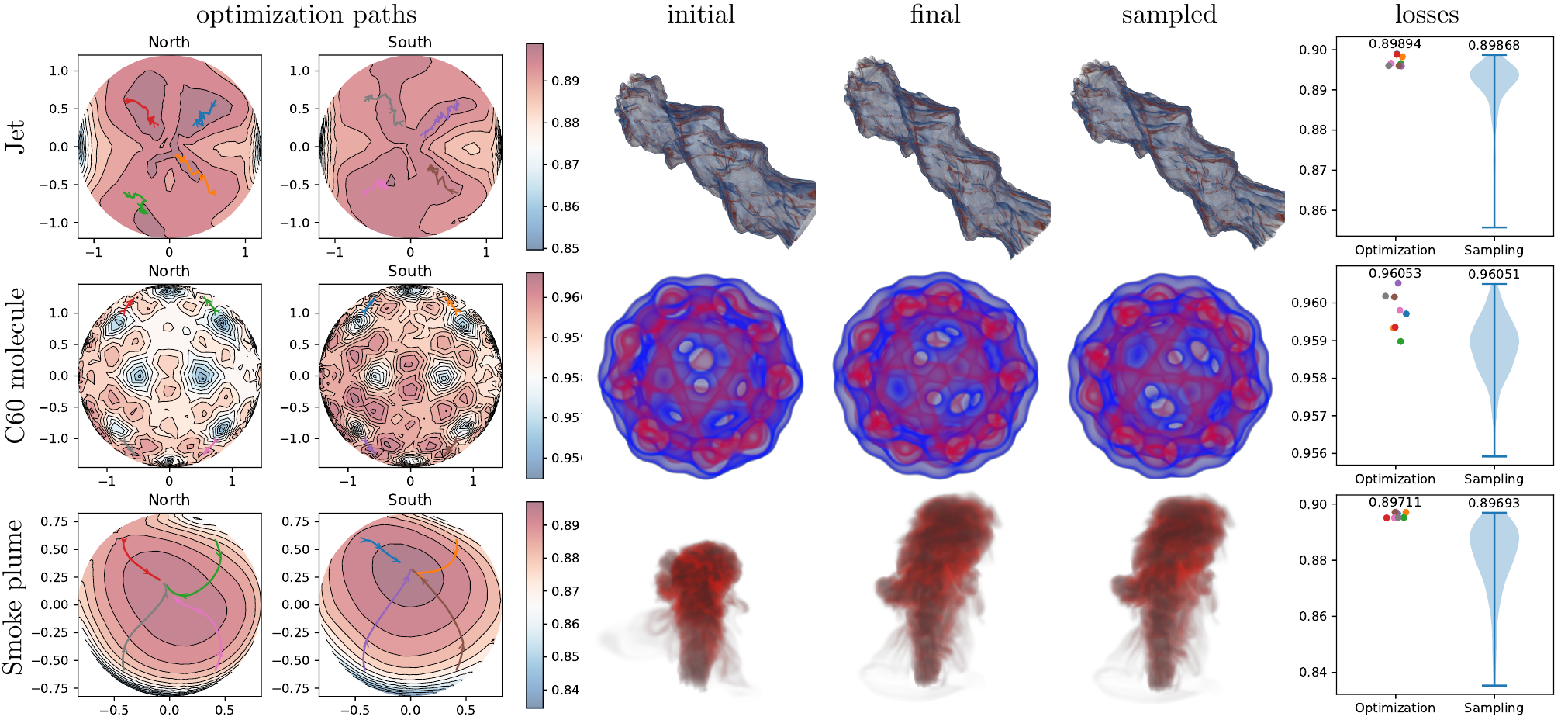}\\\vspace*{-2em}%
    \subfloat[]{\begin{minipage}{0.374\textwidth}\hfill\vspace{1mm}\end{minipage}}
    \subfloat[]{\begin{minipage}{0.450\textwidth}\hfill\vspace{1mm}\end{minipage}}
    \subfloat[]{\begin{minipage}{0.176\textwidth}\hfill\vspace{1mm}\end{minipage}}
    \vspace*{-1em}
    \caption{Best viewpoint selection using maximization of visual entropy for datasets jetstream ($256^3$), potential field of a C60 molecule ($128^3$), and smoke plume ($178^3$). Comparison of DiffDVR with eight initializations against random uniform sampling over the sphere of 256 views. (a) Optimization paths over the sphere. (b) Initial view, selected view of DiffDVR, best sampled view. 
    (c) Visual entropy of optimization results (colored points corresponding to (a)) vs. sampled images. Violin plot shows the distribution of loss values when sampling the sphere uniformly. Visual entropy of the best viewport is shown above each plot.
    }
    \label{fig:app:camera:more}
\end{figure*}

We assume that the camera is placed on a sphere enclosing the volume and faces toward the object center. The camera is parameterized by longitude and latitude. 
AD is used to optimize the camera parameters to determine the viewpoint that maximized the selected cost function.
As cost function, we adopt the differentiable opacity entropy proposed by Ji~\etAl~\cite{ji2006dynamic}.

Let $C \in \R^{H \times W \times 4}$ be the output image. We employ array notation, i.e., $C[x,y,c]$ indicates color channel $c$ (red, green, blue, alpha) at pixel $x,y$.
The entropy of a vector $\bm{x}\in\R^N$ is defined as
\begin{equation}
    H(\bm{x}) = \frac{1}{\log_2 N} \sum_{i=1}^N{p_i \log_2 p_i} \ , \ \ p_i = \frac{\bm{x}_i}{\sum_{j=1}^N{\bm{x}_j}} .
\end{equation}
Then the opacity entropy is defined as
$
    \text{OE}(C) = H(C[:,:,3]),
$
where $C[:,:,3]$ indicates the linearization of the alpha channel, and the color information is unused.

In a first experiment, the best viewpoint is computed for a CT scan of a human tooth of resolution $256 \times 256 \times 161$. Eight optimizations are started in parallel with initial views from viewpoints at a longitude of $\{45^{\circ}, 135^{\circ}, 225^{\circ}, 315^{\circ}\}$ and a latitude of $\pm 45^{\circ}$. In all cases, 20 iterations using gradient descent are  performed. The viewpoints selected by the optimizer are shown as paths over the sphere in \autoref{fig:app:camera:tooth2d}a. The values of the cost function over the course of optimization are given in \autoref{fig:app:camera:tooth2d}b. 
It can be seen that the eight optimization runs converge to three distinct local minima. The best run converges to approximately the same entropy as obtained when the best view from 256 uniformly sampled views over the enclosing sphere is taken. \autoref{fig:teaser}a shows intermediate views and the view from the optimized viewpoint. Further results on other datasets, i.e, a jetstream simulation ($256^3$), the potential field of a C60 molecule ($128^3$), and a smoke plume ($178^3$), confirm the dependency of the optimization process on the initial view (see \autoref{fig:app:camera:more}). 

Both the adjoint and the forward method compute exactly the same gradients, except for rounding errors. As seen in \autoref{fig:app:camera:tooth2d}c, a single forward/backward pass in the adjoint method requires about $9.5$ms/$14.6$ms, respectively, giving a total of $~24.1$ms.
For the forward method, we compare two alternatives. First, \emph{forward-immediate} directly evaluates the forward variables during the forward pass in PyTorch and stores these variables for reuse in the backward pass. In \emph{forward-delayed}, the evaluation of gradients is delayed until the backward pass, requiring to re-trace the volume.
With $21.3$ms, \emph{forward-immediate} is slightly faster than the adjoint method, while \emph{forward-delayed} is around $30\%$ slower due to the re-trace operation. 

%

\subsection{Transfer Function Reconstruction}\label{sec:app:tf}

Our second use case is TF reconstruction. Reference images of a volume are first rendered from multiple views using a target TF. Given the same volume and an arbitrary initial TF, AD is then used to optimize the TF so that the rendered images match the references. 
The target TF comprises of 256 rgb$\alpha$ entries, the target TF with $R$ entries is initialized with random Gaussian noise. The density volume is rendered to a $512^2$ viewport from eight camera positions that are uniformly distributed around the volume (the view direction always pointing toward the volume's center).

\begin{figure}
    \centering
    \includegraphics[width=\linewidth]{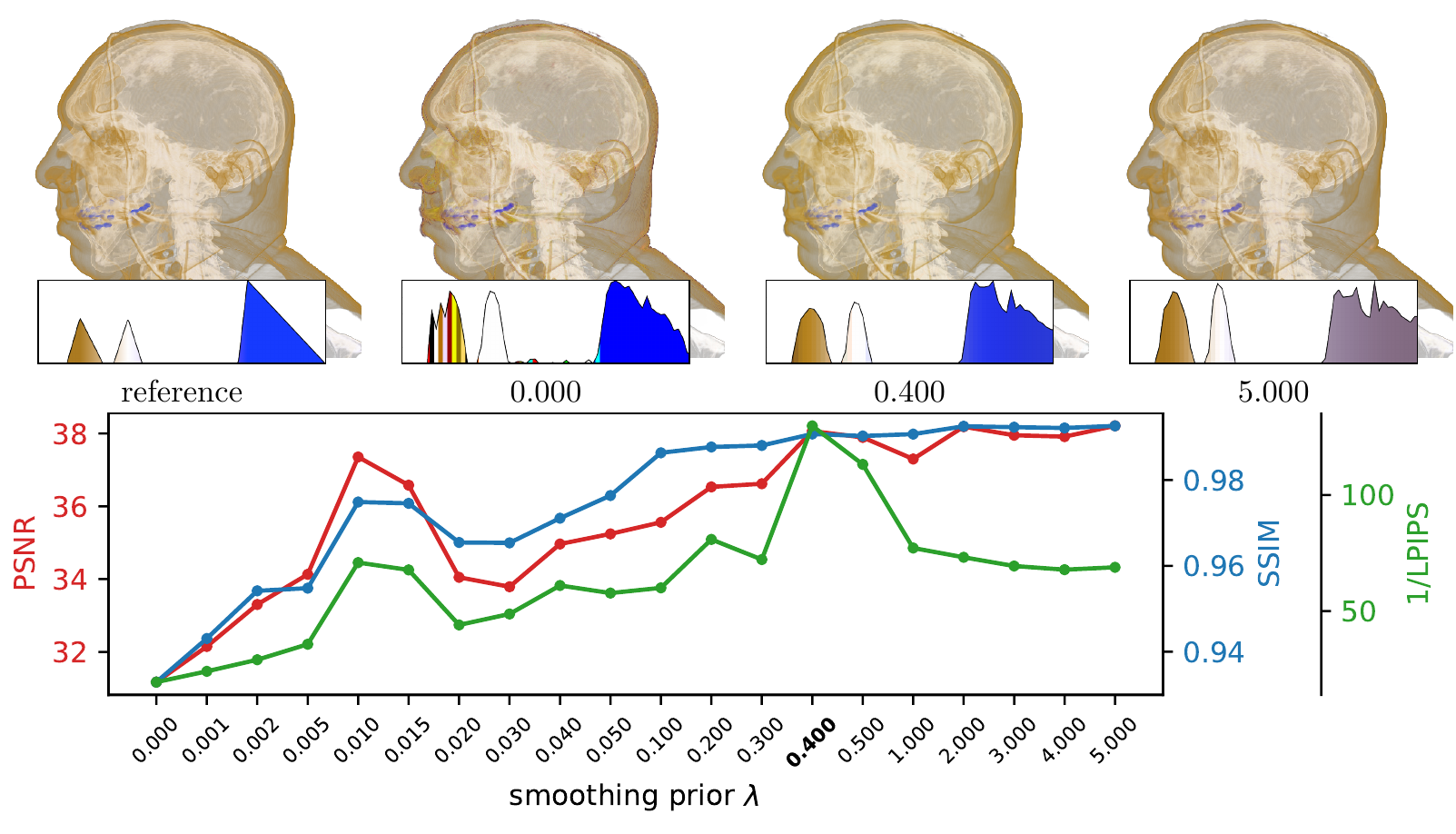}\vspace*{-1em}%
    \caption{Effect of the smoothing prior (\autoref{eq:tf-loss}). A small value of $\lambda$ leads to ``jagged'' TFs which can accurately predict small details like the teeth in blue but introduce low frequency color shifts resulting in low PSNR and SSIM~\cite{wang2004image}. A large smoothing prior smooths out small details. 
    }
    \label{fig:app:tf:smoothing}
\end{figure}

Let $T\in\R^{R,4}$ be the TF with $R$ entries containing the rgb color and absorption, and let $\bm{x}_i$ be the $N$ rendered image of resolution $W \times H$, $\bm{y}_i$ are the reference images. In our case, $N=8, W=H=512$ and $R$ varies.
We employ an $L_1$ loss on the images and a smoothing prior $L_\text{prior}$ on the TF, i.e., 
\begin{equation}
    \begin{aligned}
     L_\text{total} &= L_1(\bm{x}) + \lambda L_\text{prior}(T), \\
     L_1(\bm{x}) &= \frac{1}{NWH} \sum_{i,x,y}{|\bm{x}_{ixy}-\bm{y}_{ixy}|} \\
    L_\text{prior}(T) &= \frac{1}{4(R-1)} \sum_{c=1}^4 \sum_{r=1}^{R-1}{(T_{c,r+1}-T_{c,r})^2} .\\
       \end{aligned}
    \label{eq:tf-loss}
\end{equation}
The Adam optimizer~\cite{kingma2014adam} is used with a learning rate of $0.8$ for $200$ epochs.
The use of $\lambda$ to control the strength of the smoothing prior is demonstrated in \autoref{fig:app:tf:smoothing} for a human head CT scan as test dataset using $R=64$. 
If $\lambda$ is too small, the reconstructed TF contains high frequencies and introduces subtle color shifts over the whole image. If the smoothing prior is too large, small details are lost. We found that a value of $\lambda$ around $0.4$ leads to the best results, visually and using the Learned Perceptual Image Patch Similarity metric (LPIPS)~\cite{zhang2018perceptual}, and is thus used in our experiments. 
We chose the LPIPS metric as we found that it can accurately distinguish the perceptually best results when the peak-signal-to-noise ratio (PSNR) and the structural similarity index (SSIM)~\cite{wang2004image} result in similar scores.
The initialization of the reconstruction and the final result for a human head dataset are shown in \autoref{fig:teaser}b.

\begin{figure}
    \centering%
    \includegraphics[width=\linewidth]{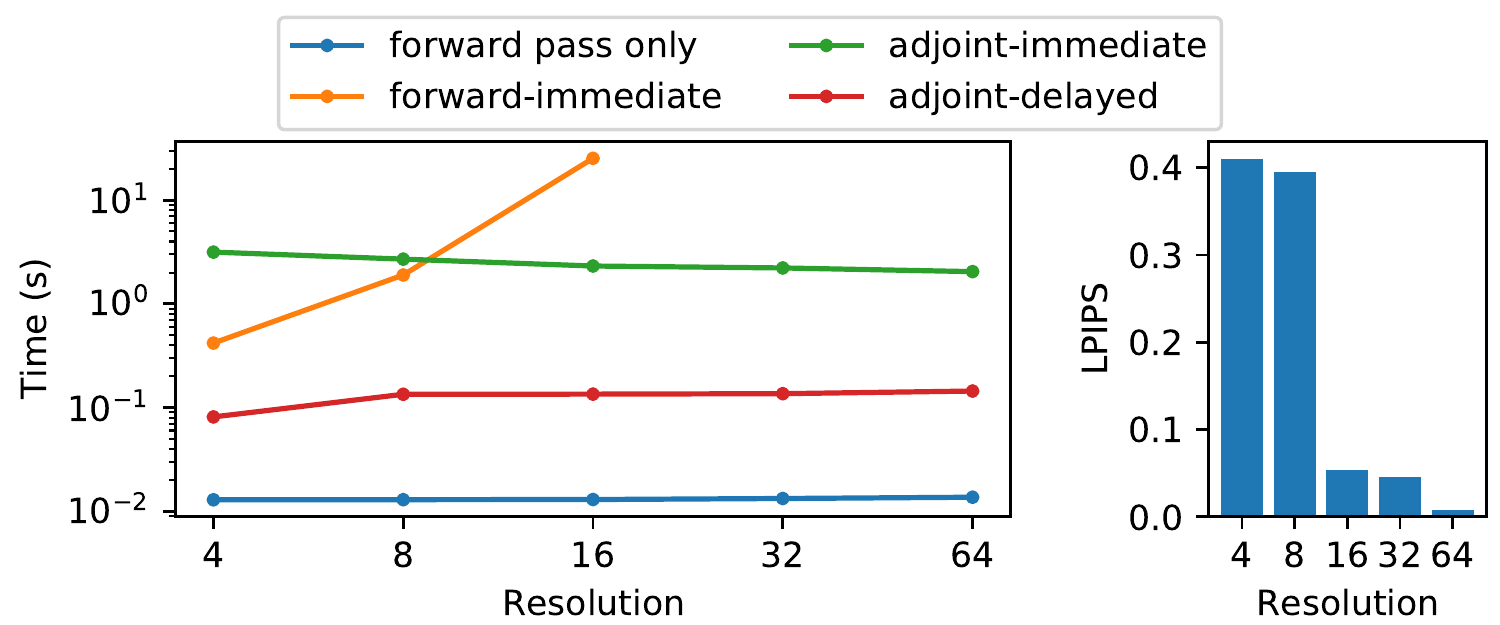}\vspace*{-1em}%
    \caption{Timings and loss function values for different AD modes and resolutions of the reconstructed TF. Timings are with respect to a single epoch.}
    \label{fig:app:tf:timings}
\end{figure}
Next, we analyze the impact of the TF resolution $R$ on reconstruction quality and performance (see \autoref{fig:app:tf:timings}).
For TF reconstruction, the backward AD mode significantly outperforms the forward mode.
Because of the large number of parameters, especially when increasing the resolution of the TF, the derivatives of many parameters have to be traced in every operation when using the forward AD mode. Furthermore, the forward variables may no longer fit into registers and overflow into global memory. This introduces a large memory overhead that leads to a performance decrease that is even worse than the expected linear decrease.
A na\"ive implementation of the adjoint method that directly accumulates the gradients for the TF in global memory using atomics is over $100\times$ slower than the non-adjoint forward pass (\emph{adjoint-immediate}). This is because of the large number of memory accesses and write conflicts.
Therefore, we employ delayed accumulation (\emph{adjoint-delayed}). The gradients for the TF are first accumulated in shared memory. Then, after all threads have finished their assigned rays, the gradients are reduced in parallel and then accumulated into global memory using atomics. As seen in \autoref{fig:app:tf:timings}, this is the fastest of the presented methods. The whole optimization for $200$ epochs requires around 5 minutes including I/O. However, as only 48kB of shared memory are freely available per multiprocessor, the maximal resolution of the TF is 96 texels. If a higher resolution is required, \emph{adjoint-immediate} must be employed.
At smaller values of $R$ the reconstruction quality is decreased (see \autoref{fig:app:tf:timings}). We found that a resolution of $R=64$ leads to the best compromise between reconstruction performance and computation time.

\begin{figure}
    \centering
    \setlength\tabcolsep{1pt}
	\setlength{\fboxsep}{0pt}%
	\newlength{\tfcompwidth}
	\setlength{\tfcompwidth}{0.23\linewidth}
	\begin{tabular}{ccccc}
	& initial & final & reference & difference (10x) \\
	    \rotatebox[origin=c]{90}{Thorax} &
		\raisebox{-0.5\height}{\begin{tikzpicture}%
			\node[inner sep=0pt] at (0,0) {\includegraphics[width=\tfcompwidth,trim=0 0 0 60,clip]{{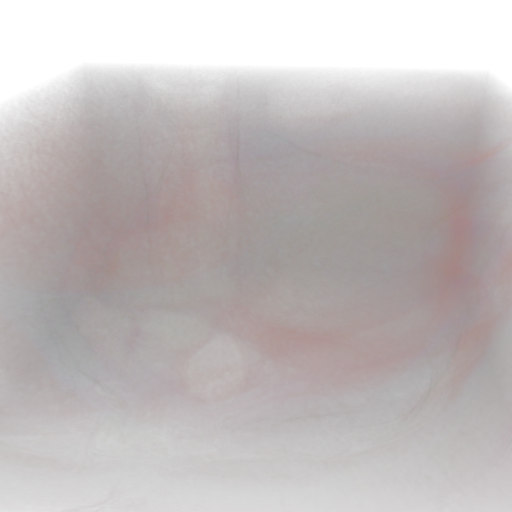}}};
			\node[inner sep=0pt,fill=white] at (0,-0.35\tfcompwidth) {\fbox{\includegraphics[width=0.8\tfcompwidth]{{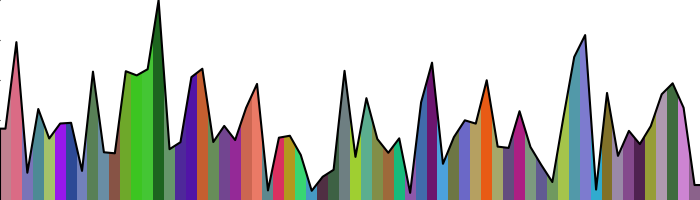}}}};
		\end{tikzpicture}}&
		\raisebox{-0.5\height}{\begin{tikzpicture}%
			\node[inner sep=0pt] at (0,0) {\includegraphics[width=\tfcompwidth,trim=0 0 0 60,clip]{{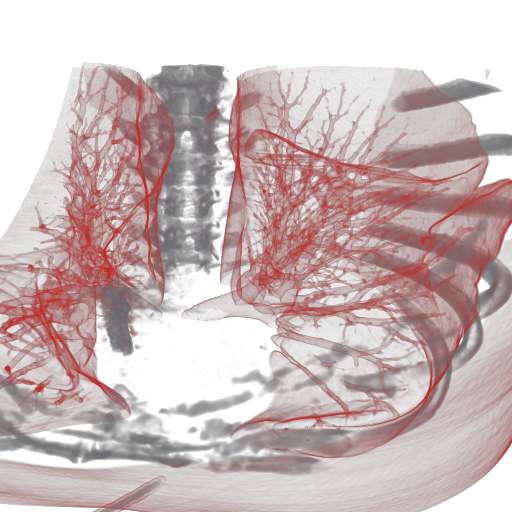}}};
			\node[inner sep=0pt,fill=white] at (0,-0.35\tfcompwidth) {\fbox{\includegraphics[width=0.8\tfcompwidth]{{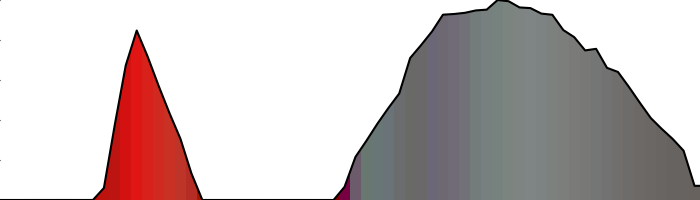}}}};
		\end{tikzpicture}}&
		\raisebox{-0.5\height}{\begin{tikzpicture}%
			\node[inner sep=0pt] at (0,0) {\includegraphics[width=\tfcompwidth,trim=0 0 0 60,clip]{{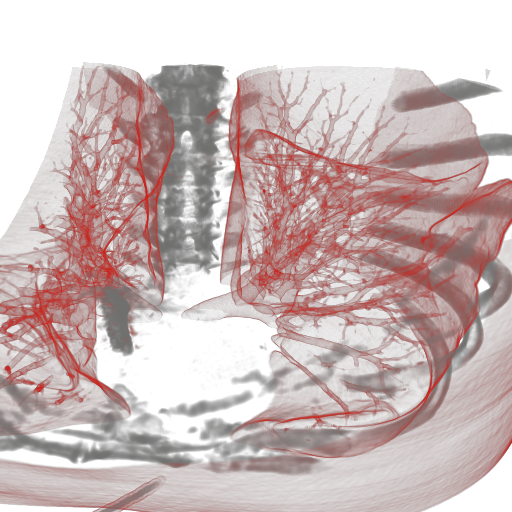}}};
			\node[inner sep=0pt,fill=white] at (0,-0.35\tfcompwidth) {\fbox{\includegraphics[width=0.8\tfcompwidth]{{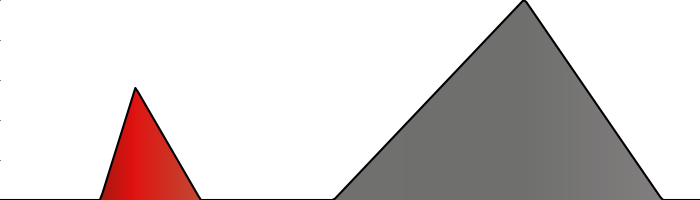}}}};
		\end{tikzpicture}}&
		\raisebox{-0.5\height}{\includegraphics[width=\tfcompwidth]{{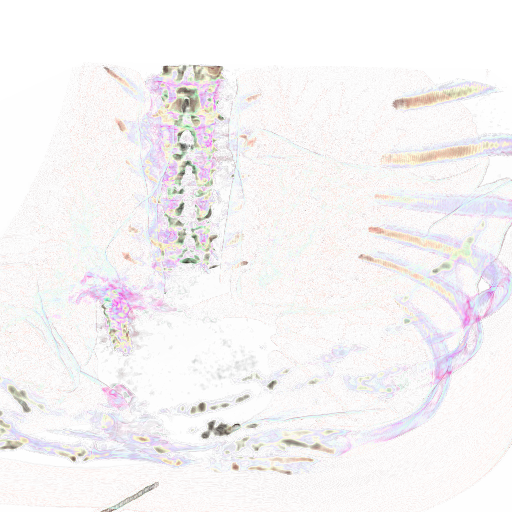}}}\\[1em]
		\rotatebox[origin=c]{90}{Plume} &
		\raisebox{-0.5\height}{\begin{tikzpicture}%
			\node[inner sep=0pt] at (0,0) {\includegraphics[width=\tfcompwidth]{{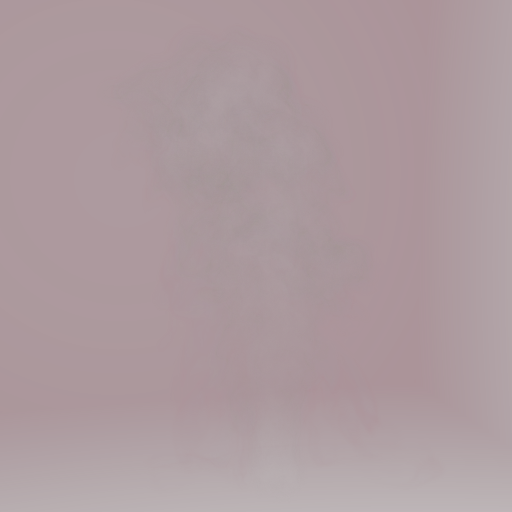}}};
			\node[inner sep=0pt,fill=white] at (0,-0.4\tfcompwidth) {\fbox{\includegraphics[width=0.8\tfcompwidth]{{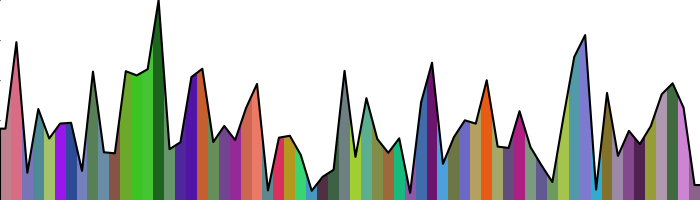}}}};
		\end{tikzpicture}}&
		\raisebox{-0.5\height}{\begin{tikzpicture}%
			\node[inner sep=0pt] at (0,0) {\includegraphics[width=\tfcompwidth]{{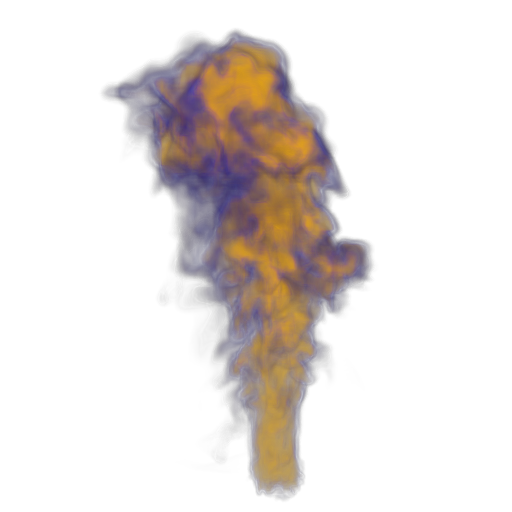}}};
			\node[inner sep=0pt,fill=white] at (0,-0.4\tfcompwidth) {\fbox{\includegraphics[width=0.8\tfcompwidth]{{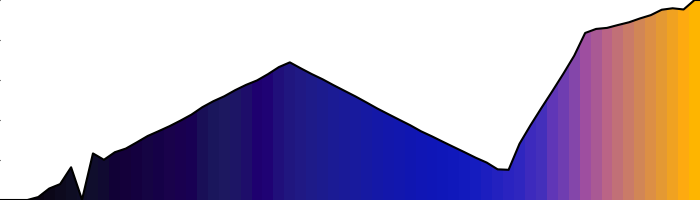}}}};
		\end{tikzpicture}}&
		\raisebox{-0.5\height}{\begin{tikzpicture}%
			\node[inner sep=0pt] at (0,0) {\includegraphics[width=\tfcompwidth]{{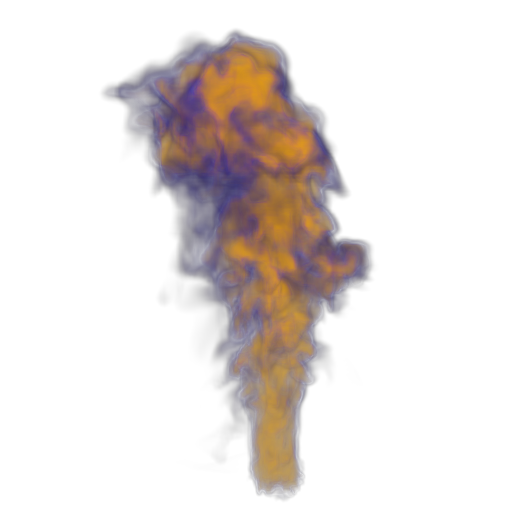}}};
			\node[inner sep=0pt,fill=white] at (0,-0.4\tfcompwidth) {\fbox{\includegraphics[width=0.8\tfcompwidth]{{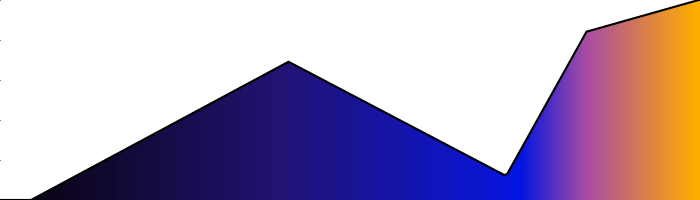}}}};
		\end{tikzpicture}}&
		\raisebox{-0.5\height}{\includegraphics[width=\tfcompwidth]{{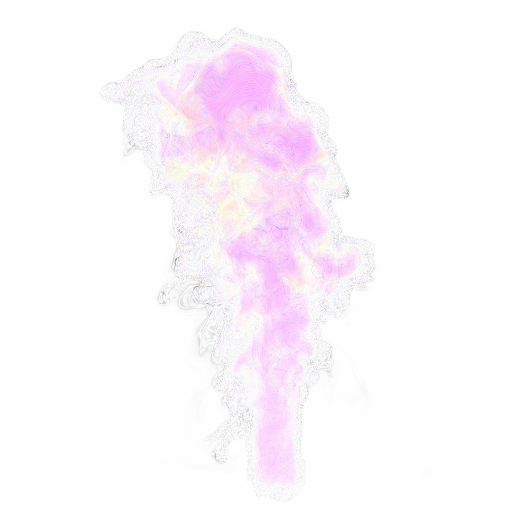}}}\\
	\end{tabular}
    \caption{TF reconstruction using a CT scan of a human thorax (PSNR=$42.6$dB) and a smoke plume (PSNR=$47.8$dB, SSIM=$0.999$). The used hyperparameters are the same as for the skull dataset. From left to right: Start configurations for the optimizer, optimized results, ground truths, pixel differences (scaled by a factor of $10$ for better perception) between ground truths and optimized results.}
    \label{fig:app:tf:more}
\end{figure}
To evaluate the capabilities of TF reconstruction to generalize to new datasets with the same hyperparameters as described above, we run the optimization on two new datasets, a CT scan of a human thorax and a smoke plume, both of resolution $256^3$. As one can see in \autoref{fig:app:tf:more}, the renderings with the reconstructed TF closely match the reference, demonstrating stability of the optimization for other datasets.

We envision that TF optimization with respect to losses in screen space can be used to generate ``good'' TFs for a dataset for which no TF is available. While a lot of research has been conducted on measuring the image quality for viewpoint selection, quality metrics specialized for TFs are still an open question to the best of our knowledge. In future work, a first approach would be to take renderings of other datasets with a TF designed by experts and transform the ``style'' of that rendering to a new dataset via the style loss by Gatys~\etAl~\cite{gatys2016image}.

\subsection{Density Reconstruction}\label{sec:app:volume}


\begin{figure*}[t]
    \centering
    \includegraphics[width=\textwidth]{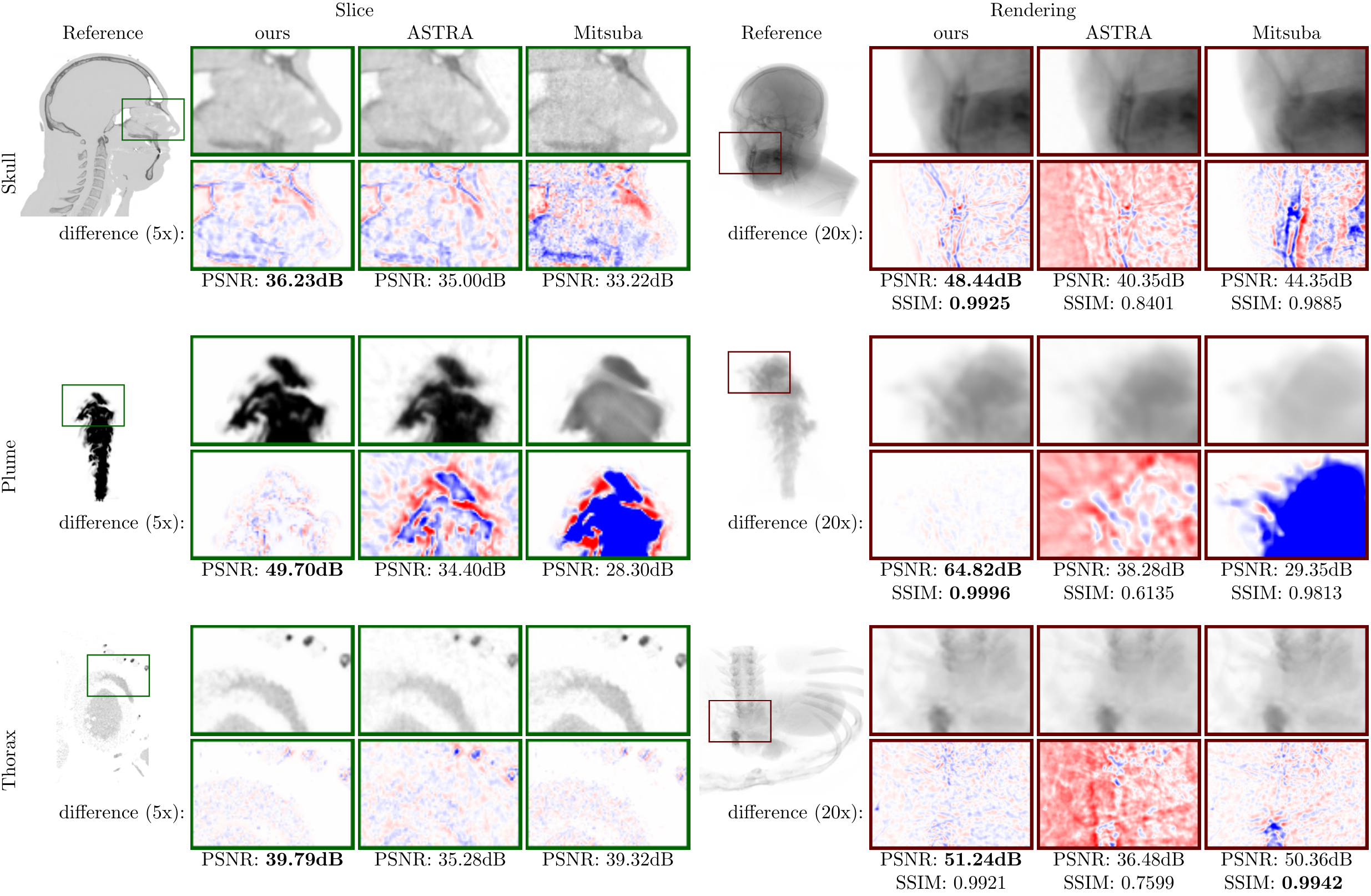}
    \caption{Density reconstruction using an optical absorption-only model. Comparison between DiffDVR, algebraic reconstruction provided by the ASTRA-toolbox~\cite{vanAarle2015astra,vanAarle2016astra} and Mitsuba's differentiable path tracer~\cite{NimierDavid2020Radiative}. For each algorithm, a single slice through the center of the reconstructed volume and a volume rendering of this volume are shown, including per-pixel differences to the reference images. PSNR values in column ``slice'' are computed over the whole volume, in column ``rendering'' they are with respect to the rendered images. 
    Timings are given in \autoref{sec:app:volume}. In the difference images, blue and red indicate under- and over-estimation, respectively.}
    \label{fig:density:linear}
\end{figure*}

In the following, we shed light on the use of DiffDVR for reconstructing a 3D density field from images of this field. For pure absorption models, the problem reduces to a linear optimization problem.
This allows for comparisons with specialized methods, such as filtered backpropagation or algebraic reconstruction~\cite{GORDON1970471,dudgeon1984multidimensional,herman2009fundamentals}.
We compare DiffDVR to the CUDA implementation of the SIRT algebraic reconstruction algorithm~\cite{gregor2008sirt} provided by the ASTRA-toolbox~\cite{vanAarle2015astra,vanAarle2016astra}. 
Furthermore, we compare the results to those computed by Mitsuba~2~\cite{NimierDavid2020Radiative,Mitsuba2}, a general differentiable path tracer.
Density reconstruction uses 64 uniformly sampled views on a surrounding sphere. Each image is rendered at a resolution of $512^2$. The reconstructed volume has a resolution of $256^3$. ASTRA and Mitsuba are used with their default optimization settings. DiffDVR performs a stepsize of $0.2$ voxels during reconstruction. The Adam optimizer with a batch size of 8 images and a learning rate of $0.3$ is used.
To speed up convergence, we start with a volume of resolution $32^3$ and double the resolution in each dimension after $10$ iterations. At the highest resolution, the optimization is performed for $50$ iterations.
The same $L_\text{total}$ loss function as for TF reconstruction (see \autoref{eq:tf-loss}) is used, except that the smoothing prior is computed on the reconstructed volume densities in 3D, with $\lambda=0.5$.

Three experiments with datasets exhibiting different characteristics are carried out. The results are shown in 
\autoref{fig:density:linear}.
As one can see, DiffDVR consistently outperforms algebraic reconstruction via ASTRA and density reconstruction via the Mitsuba framework. In particular, Mitsuba suffers from noise in the volume due to the use of stochastic path tracing. Only for the rendering of the thorax dataset, Mitsuba shows a slightly better SSIM score than DiffDVR.
For the plume dataset, intermediate results of the optimization process until convergence are shown in \autoref{fig:teaser}c. 

Note that all compared algorithms serve different purposes. Algebraic reconstruction methods (ASTRA) are specialized for absorption-only optical models and support only such models. Mitsuba is tailored to Monte Carlo path tracing with volumetric scattering, an inherently computational expensive task. DifffDVR is specialized for direct volume rendering with an emission-absorption model and a TF, yet emissions and a TF were disabled in the current experiments.
These differences clearly reflect in the reconstruction times. For instance, for reconstructing the human skull dataset, ASTRA requires only 53 seconds, DiffDVR requires around 12 minutes, and Mitsuba runs for multiple hours. 

\subsection{Color Reconstruction}\label{sec:app:color}

\begin{figure*}[t]
    \centering
    \includegraphics[width=\textwidth]{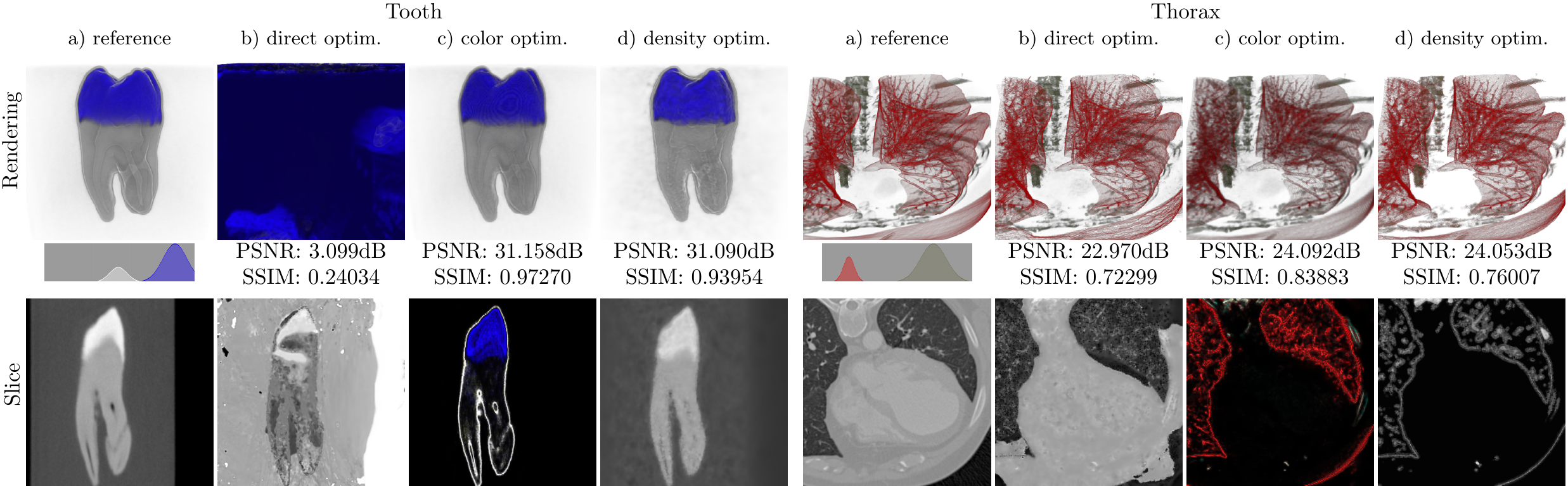}
    \caption{Density optimization for a volume colored via a non-monotonic rgb$\alpha$-TF using an emission-absorption model. (a) Rendering of the reference volume of a human tooth and a human thorax. (b) Local minimum of the loss function. (c) Pre-shaded color volume as initialization. (d) Final result of the density volume optimization with TF mapping. The second row shows slices through the volumes. Note the colored slice through the pre-shaded color volume in (c).}
    \label{fig:volume:prerecon}
\end{figure*}

\begin{figure}
    \includegraphics[width=\linewidth]{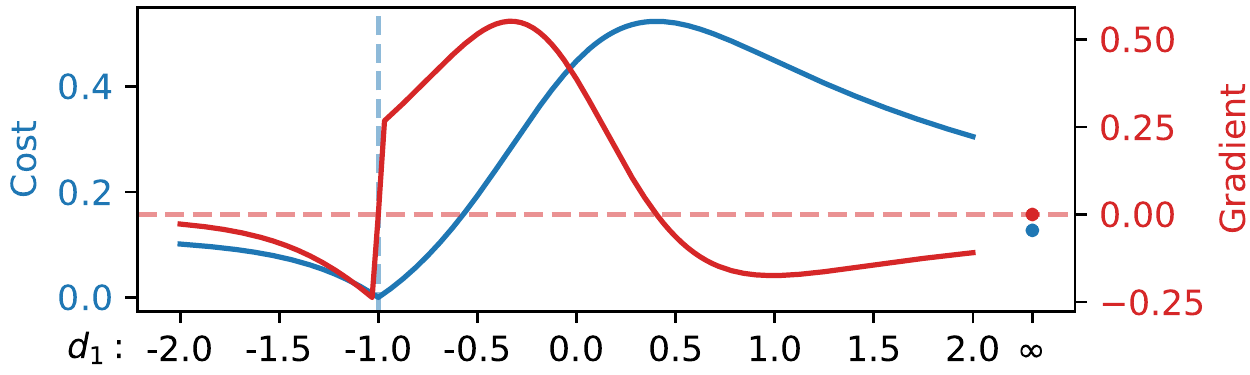}
    \caption{1D example for a density optimization with a Gaussian TF with the optimum at a density of $-1.0$. For a value $>0.4$, the gradient faces away from the optimum.}
    \label{fig:volume:gaussian}
\end{figure}

Next, we consider an optical emission-absorption model with a TF that maps densities to colors and opacities, as it is commonly used in DVR. To the best of our knowledge, we are the first to support such a model in tomographic reconstruction.

For TFs that are not a monotonic ramp, as in the absorption-only case, density optimization becomes a non-convex problem. Therefore, the optimization can be guided into local minima by a poor initialization. We illustrate this problem in a simple 1D example.
A single unknown density value $d_1$ of a 1D ``voxel'' -- a line segment with two values $d_0=-1$ and $d_1$ at the end points and linear interpolation in between -- should be optimized. A single Gaussian function with zero mean and variance $0.5$ is used as TF, and the ground truth value for $d_1$ is $-1$. For varying $d_1$, \autoref{fig:volume:gaussian} shows the $L_2$-loss between the color obtained from $d_1$ and the ground truth, and the corresponding gradients. As can be seen, for initial values of $d_1>0.4$ the gradient points away from the true solution. Thus, the optimization ``gets stuck'' at the other side of the Gaussian, never reaching the target density of $-1$.
This issue worsens in 2D and 3D, as the optimizer needs to reconstruct a globally consistent density field considering many local constraints.
This failure case is also shown in \autoref{fig:volume:prerecon}b, where the tooth dataset cannot be reconstructed faithfully due to the initialization with a poorly matching initial field.

To overcome this shortcoming, it is crucial to start the optimization with an initial guess that is not ``too far'' from the ground truth in the high-dimensional parameter space.
We account for this by proposing the following optimization pipeline:
First, a pre-shaded color volume of resolution $256^3$ (\autoref{fig:volume:prerecon}c) is reconstructed from images using the same multi-resolution optimization as in the case of an absorption-only model. The color volume stores the rgb-emission and scalar absorption per voxel, instead of a scalar density value that is mapped to color via a TF. By using this color volume, trapping into local minima with non-monotonic TFs can be avoid. Intermediate results of the optimization process until convergence are shown in \autoref{fig:teaser}d for the tooth dataset.
Then, density values that match the reconstructed colors after applying the TF are estimated. For each voxel, $256$ random values are sampled, converted to color via the TF, and the best match is chosen. To avoid inconsistencies between neighboring voxels, an additional loss term penalises differences to neighbors. Let $(\tau_T, C_T)$ be the target color from the color volume and $d$ the sampled density with mapped color $\tau(d), C(d)$, then the cost function is
\begin{equation}
    \mathcal{C}(d) = ||C_T-C(d)||_2^2 + \alpha \log(1+|\tau_T-\tau(d)|) + \beta \sum_{i\in\mathcal{N}}{(d-d_i)^2} .
\end{equation}
Here, $\alpha$ and $\beta$ are weights, and $\mathcal{N}$ loops over the 6-neighborhood of the current voxel. The logarithm accounts for the vastly different scales of the absorption, similar to an inverse of the transparency integral \autoref{eq:integral2}.
In the example, we set $\alpha=1/\max(\tau_T)$ to normalize for the maximal absorption in the color volume, and $\beta=1$.
This process is repeated until the changes between subsequent iterations fall below a certain threshold, or a prescribed number of iterations have been performed. 

Finally, the estimated density volume is used as initialization for the optimization of the density volume from the rendered images (\autoref{fig:volume:prerecon}d). We employ the same loss $L_\text{total}$ as before with a smoothing prior of $\lambda=20$.
The total runtime for a $256^3$ volume is roughly 50 minutes. Even though the proposed initialization overcomes to a certain extent the problem of non-convexity and yields reasonable results, \autoref{fig:volume:prerecon} indicates that some fine details are lost and spurious noise remains. We attribute this to remaining ambiguities in the sampling of densities from colors that still lead to suboptimal minima in the reconstruction.
This also shows in the slice view of \autoref{fig:volume:prerecon}d, especially for the thorax dataset. Here, some areas that are fully transparent due to the TF are arbitrarily mapped to a density value of zero, while the reference has a density around 0.5 -- between the peaks of the TF -- in these areas.

\section{Conclusion}\label{sec:conclusion}

In this work, we have introduced a framework for differentiable direct volume rendering (DiffDVR), and we have demonstrated its use in a number of different tasks related to data visualization. 
We have shown that differentiability of the direct volume rendering process with respect to the viewpoint position, the TF, and the volume densities is feasible, and can be performed at reasonable memory requirements and surprisingly good performance. 

Our results indicate the potential of the proposed framework to automatically determine optimal parameter combinations regarding different loss functions. This makes DiffDVR in particular interesting in combination with neural networks. Such networks might be used as loss functions -- providing blackboxes, which steer DiffDVR to an optimal output for training purposes, e.g., to synthesize volume-rendered imagery for transfer learning tasks. 
Furthermore, derivatives with respect to the volume from rendered images promise the application to scene representation networks trained in screen space instead of from points in object space. We see this as one of the most interesting future works, spawning future research towards the development of techniques that can convert large data to a compact representation -– a code -– that can be permanently stored and accessed by a network-based visualization.
Besides neural networks, we imagine possible applications in the development of lossy compression algorithms, e.g. via wavelets, where the compression rate is not determined by losses in world space, but by the quality of rendered images. The question we will address in the future is how to generate such (visualization-)task-dependent codes that can be intertwined with differentiable renderers. 

\section*{Acknowledgments}
The authors wish to thank Jakob Wenzel and Merlin Nimier-David for their help and valuable suggestions on the Mitsuba~2 framework.%


\bibliographystyle{abbrv-doi}

\bibliography{main}

\end{document}